\documentclass[12pt,hidelinks]{article}
\pdfoutput=1
\usepackage{mathrsfs,amsmath,epsfig,amssymb,color,bm}
\usepackage{algorithm,algorithmic,natbib,subcaption,hyperref,float}
\usepackage[margin=1in]{geometry}
\usepackage[symbol]{footmisc}
\usepackage{booktabs}
\usepackage{threeparttable}
\usepackage{graphicx}
\graphicspath{{figures/}}
\allowdisplaybreaks

\renewcommand{\theequation}{\arabic{section}.\arabic{equation}}
\newtheorem{theorem}{Theorem}%[section]
\newcommand{\balpha}{{\boldsymbol \alpha}}
\newcommand{\bbeta}{{\boldsymbol \beta}}

\newcommand{\black}{\color{black}}

\begin{document}

\centerline {\Large\bf A Framework for Mediation Analysis with Massive Data }
\vspace*{0.2in}

\centerline{ {\large Haixiang Zhang$^{*}$ and Xin  Li}}
 \vspace*{0.1in}

\centerline{\it \small Center for Applied Mathematics, Tianjin University, Tianjin 300072, China}

\footnotetext[1]{Corresponding author. Email: haixiang.zhang@tju.edu.cn (Haixiang Zhang)}
\vspace{1cm}

\begin{abstract}
During the past few years, mediation analysis has gained increasing popularity across various research fields. The primary objective of mediation analysis is to examine the direct impact of exposure on outcome, as well as the indirect effects that occur along the pathways from exposure to outcome. There has been a great number of articles that applied mediation analysis to data from hundreds or thousands of individuals.  With the rapid development of technology,  the volume of avaliable data increases exponentially, which brings new challenges to researchers. Directly conducting statistical analysis for large datasets is often computationally infeasible. Nonetheless, there is a paucity of findings regarding mediation analysis in the context of big data. In this paper, we propose utilizing subsampled double bootstrap and divide-and-conquer algorithms to conduct statistical mediation analysis on large-scale datasets. {\black The proposed algorithms offer a significant enhancement in computational efficiency over traditional bootstrap confidence interval and Sobel test, while simultaneously ensuring desirable confidence interval coverage and power.} We conducted extensive numerical simulations to evaluate the performance of our method.  The practical applicability of our approach is demonstrated through two real-world data examples.

\end{abstract}

%%%%%%%%%%%%%%%%%%%%%%%%%%%%%%%%%%%%%%%%%%%%%%%%%%%%%%%%%%%%%%%%%%%%%%%%%%%%%%%%%%%%%%%%%%%%%%%%%%%%%%%%%%%%%%
%%%%%%%%%%%%%%%%%%%%%%%%%%%%%%%%%%%%%%%%%%%%%%%%%%%%%%%%%%%%%%%%%%%%%%%%%%%%%%%%%%%%%%%%%%%%%%%%%%%%%%%%%%%%%%
\section{Introduction}

Mediation analysis plays a crucial role in understanding the causal mechanism by which an independent variable X influences a dependent variable Y through an intermediate variable (mediator) M.  The topic of mediation analysis is widely researched in various fields, including but not limited to psychology, economics, epidemiology, medicine, sociology and behavioral science. \cite{baron1986moderator} laid the foundation for the development of mediation analysis in academic literature. {Afterwards, numerous papers have been published to advance the development of mediation analysis. e.g.,  mediation effects with incomplete data (\citeauthor{Lijuan-SEM-2011}, \citeyear{Lijuan-SEM-2011}; \citeauthor{Zhiyong-2013}, \citeyear{Zhiyong-2013}),  propensity score based methods (\citeauthor{Booi-MBR-2011}, \citeyear{Booi-MBR-2011}; \citeauthor{Donna-2011}, \citeyear{Donna-2011}),  longitudinal mediation models (\citeauthor{Tilmann-SEM-2011}, \citeyear{Tilmann-SEM-2011}; \citeauthor{DF-Psychometrika-2014}, \citeyear{DF-Psychometrika-2014}), quantile mediation effects \cite[]{Quantile_MBR-2014},  network mediation analysis (\citeauthor{NM-Psychometrika-2021}, \citeyear{NM-Psychometrika-2021}; \citeauthor{Network-SEM-2021}, \citeyear{Network-SEM-2021}), high-dimensional mediation effects (\citeauthor{zhang-bio-2016}, \citeyear{zhang-bio-2016}; \citeauthor{Zhang-SIM-2021}, \citeyear{Zhang-SIM-2021}; \citeauthor{Zhang-Bio-2021}, \citeyear{Zhang-Bio-2021}), Bayesian approaches for estimating mediation effects (\citeauthor{Bayesian-SEM-2021}, \citeyear{Bayesian-SEM-2021};  \citeauthor{Bayesian-MBR-2021}, \citeyear{Bayesian-MBR-2021}).  For more related results on mediation analysis, we refer to three reviewing papers by \cite{David-2007}, \cite{Preacher-2015} and \cite{zhang-review-2022}.
 }

With the advancement of technology, there has been an exponential increase in the volume of available data in recent year. When dealing with massive datasets, the computational burden becomes onerous if we directly apply traditional statistical methods. There have been recent advancements in statistical methodologies for analyzing large-scale datasets. {\black For example, \cite{chen2014split} proposed a divide-and-conquer algorithm for generalized linear models with large-scale  datasets; \cite{Sengupta2016A} introduced
a novel subsampled double bootstrap method for massive data.} However, there is a lack of research on mediation analysis using large datasets. {One recent example of studies that performed mediation analysis on large dataset is attributed to \cite{Hou-SMMR-2022}.  The authors proposed a method (PSE-MR) to identify and estimate path-specific effects of  body mass index (BMI) on cardiovascular
disease (CVD) through multiple causally ordered and non-ordered mediators (e.g., lipids) using summarized genetic data. This large dataset includes 694,649 participants from the Genetic Investigation of ANthropometric Traits (GIANT). {\black There are two primary challenges that researchers may encounter when conducting mediation analysis on large datasets: The construction of bootstrap-based confidence intervals and the performance of Sobel test require high computing resources, while computation speed is relatively low. To address this issue, we present and assess the efficiency of two techniques that have the potential to reduce the computational burden associated with conducting mediation analyses on large datasets: subsampled double bootstrap  \cite[]{Sengupta2016A} for confidence intervals and the divide-and-conquer algorithm \cite[]{chen2014split} for Sobel test. Specifically, the subsampled double bootstrap method is proposed to reduce the computational costs of traditional bootstrap-based mediation confidence intervals. Meanwhile, the divide-and-conquer approach is employed to alleviate the computational burden of the conventional Sobel test. Our aim is to efficiently conduct mediation analysis on large datasets, with a primary focus on speed, utilizing subsampled double bootstrap and divide-and-conquer methods.

}

The subsequent sections of this paper are structured as follows:  In Section \ref{sec2},  we provide a comprehensive overview of the definitions and notations for linear and logistic mediation models within the potential outcomes framework. In Section \ref{sec3-SDB},  a refined approach of subsampled double bootstrap is employed to construct confidence intervals for mediation effects.  In Section \ref{secDC-4},  we introduce a Sobel mediation testing procedure based on the divide-and-conquer approach.  In Sections \ref{sec-sim} and \ref{sec-data},  we provide simulations and real-world data examples to validate the effectiveness of our methodologies.   Section \ref{sec-7r} provides concluding remarks.

\section{Models and Notations}\label{sec2}
\setcounter{equation}{0}
\subsection{Linear Mediation Model}\label{sec2-1}
%\newcounter{0}
{
 In this section, we will review the counterfactual (or potential outcome) framework for mediation models that involve multiple continuous mediators and a continuous outcome variable. Based on the notations of counterfactuals derived from literature on causal mediation analysis \cite[]{Vanderweele-SII-2009}, Let $\mathbf{M}(x)$ denote the potential value of a d-dimensional vector of mediators, represented as $(M_{1}(x), \cdots, M_{d}(x))^\prime$, under exposure level $x$. Denote $Y(x,\mathbf{m})$ as the potential outcome when $x$ is the exposure and $\mathbf{m}=(m_1,\cdots,m_d)^\prime$ are the mediators. The linear mediation models, represented by counterfactual notations, are defined as follows:
\begin{eqnarray}
Y(x,\mathbf{m})&=&c + \gamma x + \beta_1 m_1 + \cdots + \beta_d m_d + \bm{\theta}^\prime \mathbf{Z} + \epsilon,\label{CML-1}\\
M_k(x) &=& c_k + \alpha_k x + {\boldsymbol\eta}_k^\prime \mathbf{Z} + e_k,~~~ k = 1, \cdots, d,\label{CML-2}
\end{eqnarray}
where $\mathbf{Z} = (Z_1,\cdots, Z_q)^\prime$ represents a vector of confounding variables or covariates; $\gamma$ represents the impact of exposure on the outcome while controlling for mediators and covariate; $\bbeta = (\beta_1, \cdots,\beta_d)^\prime$ represents the relationship  between mediators and outcome;
$\balpha = (\alpha_1,\cdots,\alpha_d)^\prime$ represents the association between exposure and mediators while adjusting for $\mathbf{Z}$; $\bm{\theta}$ and ${\boldsymbol\eta}_k$ denote the regression coefficients of $\mathbf{Z}$; $c$ and $c_k$ are intercepts; $\epsilon$ is mean zero normal error, $\mathbf{e}=(e_1,\cdots,e_d)^\prime$ is a mean zero normal vector with covariance matrix ${\boldsymbol\Sigma}_e$.

To decompose the effect of an exposure on outcome into direct and indirect effects, we will first review some fundamental concepts within the framework of causal mediation analysis.

$\bullet$ Natural direct effect (NDE):
\begin{eqnarray*}
{\rm NDE} &=& E[Y(x,\mathbf{M}(x^*))] - E[Y(x^*,\mathbf{M}(x^*))]\\
&=&\int_\mathbf{m} \{E(Y|x,\mathbf{m}) - E(Y|x^*,\mathbf{m}\} P(\mathbf{M} = \mathbf{m}|x^*)d\mathbf{m}.
\end{eqnarray*}

$\bullet$ Natural indirect effect (NIE):
\begin{eqnarray*}
{\rm NIE} &=& E[Y(x,\mathbf{M}(x))] - E[Y(x,\mathbf{M}(x^*))]\\
&=&\int_\mathbf{m} E(Y|x,\mathbf{m})\{P(\mathbf{M}=\mathbf{m}|x) - P(\mathbf{M}=\mathbf{m}|x^*)\}d\mathbf{m}.
\end{eqnarray*}

$\bullet$ Total effect (TE):
\begin{eqnarray*}
{\rm TE} &=& E[Y(x,\mathbf{M}(x))] - E[Y(x^*,\mathbf{M}(x^*))]\\
&=& {\rm NIE} + {\rm NDE}.
\end{eqnarray*}
For the purpose of identifying causal effects, the causal mediation analysis literature requires adherence to four fundamental assumptions \cite[]{Vanderweele-SII-2009}.
\begin{itemize}
\item[(C.1)] Stable Unit Treatment Value Assumption (SUTVA). There is no multiple versions of exposures and there is no interference  between subjects, which implies that the observed variables are identical to the potential variables corresponding to the actually observed exposure level. i.e., $\mathbf{M} = \sum_x \mathbf{M}(x)I(X=x)$, and $Y=\sum_x\sum_\mathbf{m} Y(x,\mathbf{m})I(X=x, \mathbf{M}=\mathbf{m})$, where $I(\cdot)$  is the indicator function.
\item[(C.2)]  There are no measurement errors in the mediators and the outcome.
\item[(C.3)]  Sequential ignorability: (i) $Y(x,\mathbf{m})\perp X|\mathbf{Z}$, i.e., no unmeasured confounding between exposure and the potential outcome. (ii) $Y(x,\mathbf{m})\perp \mathbf{M}|\{X,\mathbf{Z}\}$, i.e., no unmeasured confounding for the mediator-outcome relationship after adjusting for the exposure. (iii) $\mathbf{M}(x) \perp X|\mathbf{Z}$, i.e.,  no unmeasured confounding for the exposure effect on all the mediators. (iv) $Y(x,\mathbf{m})\perp \mathbf{M}(x^*)|\mathbf{Z}$, i.e., no exposure-induced confounding between mediators
and the potential outcome.
\item[(C.4)] The mediators are assumed to be causally independent. i.e., it is not allowed that one mediator is the cause of another.
\end{itemize}

\begin{theorem}\label{TH-1}
Under the assumptions (C.1)-(C.4), if the mediation models (\ref{CML-1}) and (\ref{CML-2})
are correctly specified, then we have
\begin{align*}
{\rm NDE}&= \gamma(x-x^*),\\
{\rm NIE}&=\sum_{k=1}^d \alpha_k\beta_k(x-x^*),\\
{\rm TE}&= \Big(\gamma + \sum_{k=1}^d \alpha_k\beta_k\Big)(x-x^*).
\end{align*}
\end{theorem}
The results of Theorem \ref{TH-1} can be similarly derived as equations (2.6), (2.7) and (2.8) in \cite{SongYY-PhD-2020}, thus the proof details are omitted here.
}

%{\blue Using  observed data, we also consider the structural equation modeling (SEM)  based framework for mediation models with multiple
%continuous mediators and a  continuous outcome} (see Figure \ref{fig-1}):
%\begin{eqnarray}
%Y &=& c + \gamma X + \beta_1 M_1 + \cdots + \beta_d M_d + \bm{\theta}^\prime \mathbf{Z} + \epsilon,\label{Eq2.1}\\
%M_k &=& c_k + \alpha_k X + {\boldsymbol\eta}_k^\prime \mathbf{Z} + e_k,~~~ k = 1, \cdots, d,\label{LM-2}
%\end{eqnarray}
%where $Y$ is the response (outcome), $X$ is an exposure,  $\mathbf{M} = (M_{1}, \cdots, M_{d})^\prime$ is a vector of mediators, $\mathbf{Z} = (Z_1,\cdots, Z_q)^\prime$ is a vector of confounding variables or covariates; $\gamma$ is the effect of
%$X$ on $Y$ adjusting for $\mathbf{M}$ and $\mathbf{Z}$; $\beta_k$ represents the relation between $M_k$ and $Y$ adjusting for other variables;
%$\alpha_k$ denotes the relation between $X$ and $M_k$ adjusting for $\mathbf{Z}$; other variables have the same meanings as that of models (\ref{CML-1}) and (\ref{CML-2}).
\begin{figure}[H]%Figure 1
\centering
\includegraphics[scale=0.4]{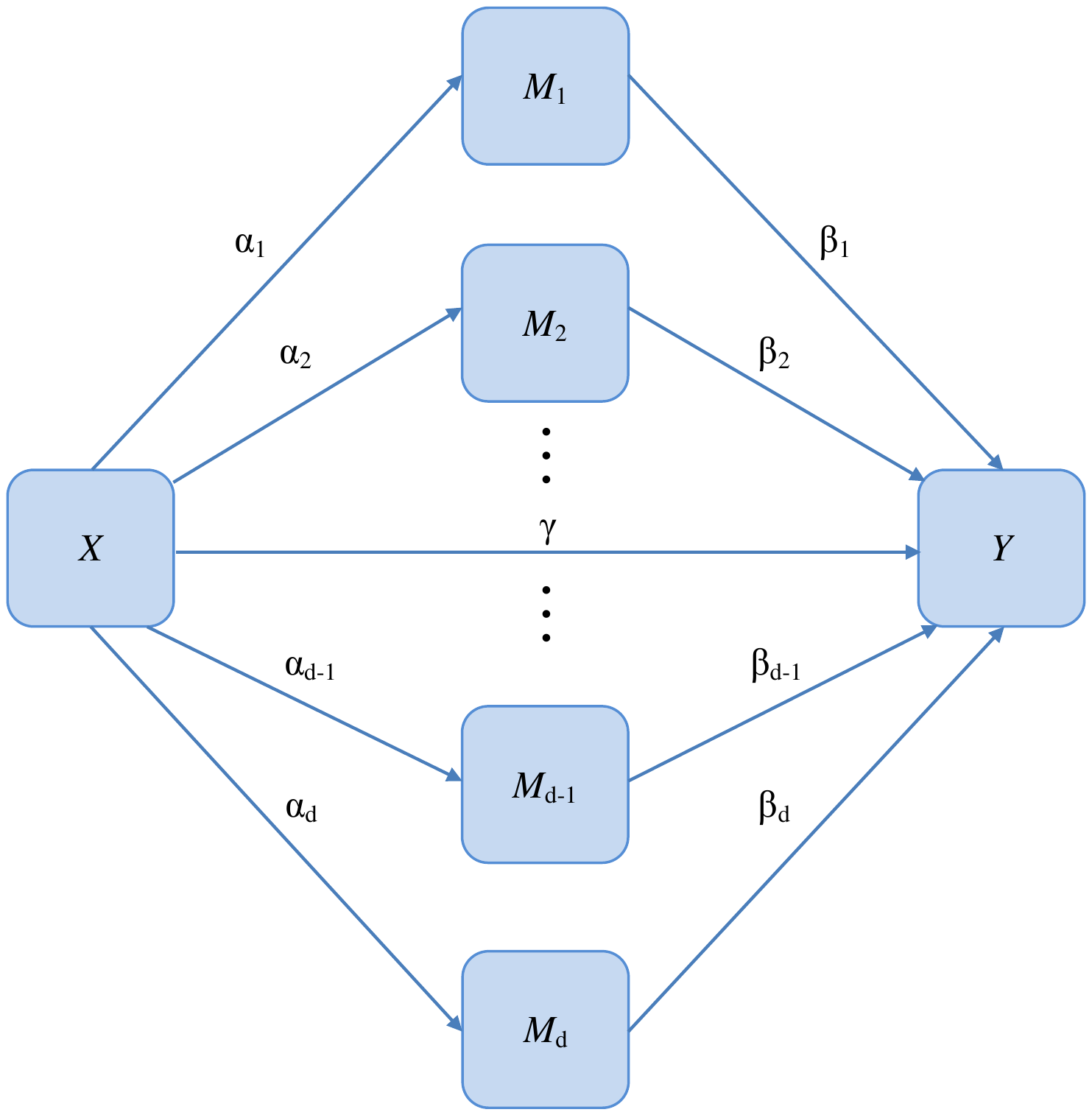}\\
\vspace{-1cm}
\begin{center}
 \caption {A scenario involving multiple mediators (excluding confounding variables).}\label{fig-1}
\end{center}
\end{figure}

{Based on Theorem \ref{TH-1} and \cite{MacKinnon-PS-2020}, we know that the total effect of $X$ on $Y$ is  $\gamma + \sum_{k=1}^d \alpha_k\beta_k $, where $\gamma$ is the {\it direct effect}, $\sum_{k=1}^d \alpha_k\beta_k $ is the {\it indirect effect}, and  $\alpha_k\beta_k$ can be interpreted as the causal indirect effect transmitted by the $k$th mediator $M_k$ along the pathway $X\rightarrow M_k\rightarrow Y$ (see Figure \ref{fig-1}) for $k=1,\cdots,d$. i.e., the product-coefficient method of SEM-based mediation framework has the same causal interpretation as  counterfactual approach when there is no X-M interaction term. }

\subsection{Logistic Mediation Model}
{
For the case of a binary outcome, we consider the counterfactual mediation models that involve continuous mediators and a binary outcome:
\begin{eqnarray}
P\{Y(x,\mathbf{m})=1\} &=& \frac{\exp\{c+ \gamma x + \beta_1 m_1 + \cdots + \beta_d m_d + \bm{\theta}^\prime \mathbf{Z}\}}{1+\exp\{c+\gamma x + \beta_1 m_1 + \cdots + \beta_d m_d + \bm{\theta}^\prime \mathbf{Z}\}},\label{CLOG-1}\\
M_k(x) &=& c_k+\alpha_k x + {\boldsymbol\eta}_k^\prime \mathbf{Z} + e_k,~~~ k = 1, \cdots, d,\label{CLOG-2}
\end{eqnarray}
where $Y(x,\mathbf{m})\in \{0,1\}$ is the potential outcome, $\mathbf{e}= (e_1,\cdots,e_d)^\prime$ is a vector of normal errors with mean zero and covariance matrix ${\boldsymbol\Sigma}_e$, other variables and parameters are similarly defined as the linear mediation models in Section \ref{sec2-1}. Following \cite{VanderWeele-AJE-2010}, we adopt the
odds ratio scale for definitions of direct and indirect effects when the outcome is rare. Based on similar counterfactual notations as  Section \ref{sec2}, the natural direct effect on the odds ratio scale ($\rm {NDE^{OR}}$) has the form
\begin{eqnarray}\label{ORNED}
\rm {NDE^{OR}} = \frac{P\{Y(x,\mathbf{M}(x^*))=1\}/[1-P\{Y(x,\mathbf{M}(x^*))=1\}]}{P\{Y(x^*,\mathbf{M}(x^*))=1\}/[1-P\{Y(x^*,\mathbf{M}(x^*))=1\}]};
\end{eqnarray}
the natural indirect effect on the odds ratio scale ($\rm {NIE^{OR}}$) is given by
\begin{eqnarray}\label{ORNIE}
\rm {NIE^{OR}} = \frac{P\{Y(x,\mathbf{M}(x))=1\}/[1-P\{Y(x,\mathbf{M}(x))=1\}]}{P\{Y(x,\mathbf{M}(x^*))=1\}/[1-P\{Y(x,\mathbf{M}(x^*))=1\}]};
\end{eqnarray}
the total effect on the odds ratio scale ($\rm {TE^{OR}}$) is
\begin{eqnarray}\label{ORTE}
\rm {TE^{OR}} = \frac{P\{Y(x,\mathbf{M}(x))=1\}/[1-P\{Y(x,\mathbf{M}(x))=1\}]}{P\{Y(x^*,\mathbf{M}(x^*))=1\}/[1-P\{Y(x^*,\mathbf{M}(x^*))=1\}]}.
\end{eqnarray}
From the above three definitions, we have this expression $\rm {TE^{OR}} = \rm {NIE^{OR}}\cdot\rm {NDE^{OR}}$.  That is  to say, $\log(\rm{TE^{OR}}) = \log(\rm {NIE^{OR}}) + \log(\rm {NDE^{OR}})$ with the log scale.

\begin{theorem}\label{Th2}
Assuming that conditions (C.1)-(C.4) are satisfied and the event is rare, then according to models (\ref{CLOG-1}) and (\ref{CLOG-2}), we can conclude that
\begin{eqnarray*}
\rm {NDE^{OR}}&=& \exp\{\gamma(x-x^*)\},\\
\rm {NIE^{OR}}&=& \exp\left\{\sum_{k=1}^d\alpha_k\beta_k(x-x^*)\right\},\\
\rm {TE^{OR}} &=&\exp\left\{\Big(\gamma + \sum_{k=1}^d\alpha_k\beta_k\Big)(x-x^*)\right\}.
\end{eqnarray*}
\end{theorem}

}

%With  observed variables, we also obtain the following SEM-based mediation model with continuous mediators and a binary outcome:
%\begin{eqnarray}
%P\{Y=1\} &=& \frac{\exp\{c+\gamma X + \beta_1 M_1 + \cdots + \beta_d M_d + \bm{\theta}^\prime \mathbf{Z}\}}{1+\exp\{c+\gamma X + \beta_1 M_1 + \cdots + \beta_d M_d + \bm{\theta}^\prime \mathbf{Z}\}},\label{EqLSE3.6}\\
%M_k &=& c_k+\alpha_k X + {\boldsymbol\eta}_k^\prime \mathbf{Z} + e_k,~~~ k = 1, \cdots, d,\label{EqLSE3.7}
%\end{eqnarray}
%where the variables and parameters are similarly defined as models (\ref{CLOG-1}) and (\ref{CLOG-2}).

Given the rarity of the outcome and assuming that (C.1)-(C.4) are satisfied, Theorem \ref{Th2} suggests that we may employ the product-of-coefficients $\alpha_k\beta_k$ to describe the mediation effect directly along the causal pathway $X\to M_k\to Y$ (see Figure 1). This stems from the conventional regression-based approach to mediation proposed by \cite{baron1986moderator}. We recommend referring to \cite{VanderWeele-AJE-2010} for a more comprehensive understanding of logistic mediation models with a single mediator.

\section{Subsampled Double Bootstrap Confidence Intervals}\label{sec3-SDB}
\setcounter{equation}{0}

The Bootstrap \cite[]{Efron-1979} is a widely used and powerful statistical tool, but its computational demands can be  extremely high for large-scale datasets.
\cite{Sengupta2016A} proposed a subsampled double bootstrap (SDB) method to deal with this problem. Its key idea is to randomly draw a small subset of the full data and use it to construct a full-size resample by generating repetition frequencies for each data point in the subset.  Although the resample has the same sample size as the original full data, it only contains a small number of unique data points. The resamples are utilized in the same manner as the conventional Bootstrap.  Notice that the computational cost of SDB mainly arises from the small subset rather than the full data set. Therefore, utilizing the SDB method significantly reduces computational costs when performing Bootstrap for massive data sets.

{\black
Suppose we have a set of $n$ independent and identically distributed samples, denoted as $\mathcal{D}_n = \{(X_i,\mathbf{M}_i,Y_i,\mathbf{Z}_i)\}_{i=1}^n$. The confidence interval for the mediation effect is crucial in understanding the statistical impact of a mediator at a specified level of reliability, such as \cite{Jeremy-MBR-2010}, \cite{preacher2012advantages} and \cite{kisbu2014distribution}. Our objective is to utilize the SDB method in the context of massive data to construct confidence intervals for the causal mediation effects $\{\alpha_k\beta_k\}_{k=1}^d$. Following \cite{Sengupta2016A}, we define the ``root function" of SDB method as
\begin{eqnarray}
T_k = \sqrt{n}(\hat\alpha_k\hat\beta_k-\alpha_k\beta_k),
\end{eqnarray}
where $\alpha_k\beta_k$ is the true mediation effect of the $k$th mediator, $\hat\alpha_k$ and $\hat\beta_k$ are the estimators of $\alpha_k$ and $\beta_k$, respectively. Let  $q^{(k)}_{v}$ represent the $v$-quantile of $T_k$, which can be estimated using the SDB procedure as follows: We first randomly select a subset of $b$ points without replacement, $\mathcal{D}^{*}_{b}=\{(X^{*}_{1},\mathbf{M}^{*}_{1},Y^{*}_{1},\mathbf{Z}^{*}_{1}),
\cdots,(X^{*}_{b},\mathbf{M}^{*}_{b},Y^{*}_{b},\mathbf{Z}^{*}_{b})\}$,  from the full data set $\mathcal{D}_n$, where  $b=n^r$ for some $0<r<1$. It is noteworthy that the elements in $\mathcal{D}^{*}_{b}$ are distinct data points.  By repeating this procedure $S$ times, we obtain $S$ subsets denoted as $\mathcal{D}^{*(s)}_{b}=\{(X^{*(s)}_{1},\mathbf{M}^{*(s)}_{1},Y^{*(s)}_{1},\mathbf{Z}^{*(s)}_{1}),
\cdots,(X^{*(s)}_{b},\mathbf{M}^{*(s)}_{b},Y^{*(s)}_{b},\mathbf{Z}^{*(s)}_{b})\}$ for $s=1,\cdots,S$. Second, we generate a resample of size $n$ denoted  by $(\mathcal{D}^{*(s)}_{b},\mathbf{W}^{*(s)}_{b})$, where $\mathbf{W}^{*(s)}_{b}=(w^{(s)}_{1},\cdots,w^{(s)}_{b})^\prime$ is a vector representing the frequencies of $\{(X^{*(s)}_{i},\mathbf{M}^{*(s)}_{i},Y^{*(s)}_{i},\mathbf{Z}^{*(s)}_{i})\}_{i=1}^b$ in the resample for $s=1,\cdots,S$. The weight vector  $\mathbf{W}^{*(s)}_{b}$ satisfies $w^{(s)}_{1}+\cdots+w^{(s)}_{b}=n$. That is to say, the vector $(w^{(s)}_{1} ,\cdots,w^{(s)}_{b})^\prime$ follows from $PN(n,\frac{1}{b},\cdots,\frac{1}{b})$, where $PN(n,\frac{1}{b},\cdots,\frac{1}{b})$ represents an $n$-trial uniform multinomial distribution over $b$ objects. Third, let $\hat\alpha_{k}^{*(s)}$ and $\hat\beta_{k}^{*(s)}$ be the estimators based on $(\mathcal{D}^{*(s)}_{b},\mathbf{W}^{*(s)}_{b})$. Calculate the values of root functions
\begin{equation}\label{Eq2.15}
T^{*(s)}_{k}= \sqrt{n} \{\hat\alpha^{*(s)}_{k}\hat\beta^{*(s)}_{k}-\hat\alpha_{k}\hat\beta_{k}\},~~~s=1,\cdots, S,
\end{equation}
where $\hat\alpha_{k}$ and $\hat\beta_{k}$ are the estimators derived from  the full data $\mathcal{D}_n$.

Denote $q^{*(k)}_{v}$ as the $v$-quantile of $\{T^{*(s)}_{k}\}_{s=1}^S$, which estimates the theoretical quantile $q^{(k)}_{v}$ of $T_k$. 
Subsequently, a single ($1-\delta$)-confidence interval for $\alpha_k\beta_k$ is  estimated as
 \begin{align}\label{Eq2.7}
{\rm CI_{single,k}}=[\hat\alpha_k\hat\beta_k - n^{-1/2} q^{*(k)}_{1-\delta/2}, ~\hat\alpha_k\hat\beta_k - n^{-1/2}q^{*(k)}_{\delta/2}],
\end{align}
where $\delta$ represents the level of significance.
Utilizing the Bonferroni technique \cite[]{Boifir-1977}, we can construct Bonferroni adjusted confidence intervals for mediators $M_k$'s as follows:
\begin{align}\label{Eq2.8}
{\rm CI_{adjusted,k}}=[\hat\alpha_k\hat\beta_k - n^{-1/2} q^{*(k)}_{1-\delta/{2d}},~\hat\alpha_k\hat\beta_k - n^{-1/2}q^{*(k)}_{\delta/{2d}}], \end{align}
where $k = 1, \cdots, d$.
These intervals satisfy $P\{\alpha_1\beta_1\in {\rm CI_{adjusted,1}}, \cdots, \alpha_d\beta_d\in {\rm CI_{adjusted,d}}\}=1-\delta$. To clarify, the confidence intervals presented in (\ref{Eq2.8}) have undergone Bonferroni correction to account for multiple testing. }

{\black We suggest employing the subsequent weighted least squares estimation equations for a linear mediation model:
\begin{equation}\label{Eq2.9}
Q^{(s)}_1(c,\gamma,\boldsymbol\beta, {\boldsymbol\theta})=\sum_{i=1}^bw^{(s)}_{i}\{Y^{*(s)}_i-c-\gamma X^{*(s)}_i - \beta_1 M^{*(s)}_{i1} - \cdots - \beta_d M^{*(s)}_{id} - \theta_1 Z^{*(s)}_{i1} - \cdots - \theta_q Z^{*(s)}_{iq}\}^2,
\end{equation}
and
\begin{equation}\label{Eq2.10}
Q^{(s)}_{2k}(c_k,\alpha_k,{\boldsymbol\eta}_k)=\sum_{i=1}^bw^{(s)}_{i}\{M^{*(s)}_{ik}-c_k-\alpha_k X^{*(s)}_i - \eta_{k1} Z^{*(s)}_{i1} - \cdots - \eta_{kq} Z^{*(s)}_{iq}\}^2, ~~~k = 1, \cdots, d,
\end{equation}
where $\boldsymbol\beta=(\beta_1,\cdots,\beta_d)^\prime$, $\boldsymbol\theta=(\theta_1,\cdots, \theta_q)^\prime$ and ${\boldsymbol\eta}_k=(\eta_{k1},\cdots,\eta_{kq})^\prime$.  The weighted least squares estimators  $\hat\beta^{*(s)}_{k}$ and $\hat\alpha^{*(s)}_{k}$ can be easily obtained by
minimizing (\ref{Eq2.9}) and (\ref{Eq2.10}), respectively, for $k=1,\cdots,d$.

For the logistic mediation model, we propose the following weighted negative log-likelihood function:
\begin{equation}\label{Q3s}
Q^{(s)}_3(c,\gamma,\boldsymbol\beta,{\boldsymbol\theta})= -\sum_{i=1}^bw^{(s)}_{i}[Y^{*(s)}_i\log p_i^{*(s)}(c,\gamma,\boldsymbol\beta,{\boldsymbol\theta}) + (1-Y^{*(s)}_i)\log\{1-p_i^{*(s)}(c,\gamma,\boldsymbol\beta,{\boldsymbol\theta})\}]^2,
\end{equation}
where
\begin{align*}
p_i^{*(s)}(c,\gamma,\boldsymbol\beta,{\boldsymbol\theta})= \frac{\exp\{c+\gamma X_i^{*(s)} +  \beta_1 M^{*(s)}_{i1} + \cdots +\beta_d M^{*(s)}_{id} + \theta_1 Z^{*(s)}_{i1} + \cdots + \theta_q Z^{*(s)}_{iq}\}}{1+\exp\{c+\gamma X_i^{*(s)} +  \beta_1 M^{*(s)}_{i1} + \cdots +\beta_d M^{*(s)}_{id} + \theta_1 Z^{*(s)}_{i1} + \cdots + \theta_q Z^{*(s)}_{iq}\}}.
\end{align*}
The corresponding estimators $\{\hat\alpha^{*(s)}_{k}\}_{k=1}^d$ and $\{\hat\beta^{*(s)}_{k}\}_{k=1}^d$ can be obtained  through the minimization of (\ref{Eq2.10}) and (\ref{Q3s}), respectively, corresponding to their respective parameters.

}
Algorithm \ref{algom1} summarizes the SDB-based method proposed in this study.
\begin{algorithm}[H]
\caption{SDB-based  Confidence Intervals for Mediation Effects}\label{algom1}
$\bullet$ {\it Step 1}: We randomly  select  a subset $\mathcal{D}^{*(s)}_{b}$ of size $b$ ($b\ll n$) without replacement from the full data $\mathcal{D}_n$.\\
$\bullet$ {\it Step 2}: Generate a resample $(\mathcal{D}^{*(s)}_{b},\mathbf{W}^{*(s)}_{b})$, where  $\mathbf{W}^{*(s)}_{b}= (w^{(s)}_{1} ,\cdots,w^{(s)}_{b})^\prime$ is a vector representing the frequencies of $\{(X^{*(s)}_{i},\mathbf{M}^{*(s)}_{i},Y^{*(s)}_{i},\mathbf{Z}^{*(s)}_{i})\}_{i=1}^b$ in the resample, Specifically,  $(w^{(s)}_{1} ,\cdots,w^{(s)}_{b}) \sim PN(n,\frac{1}{b},\cdots,\frac{1}{b})$.\\
$\bullet$ {\it Step 3}: Calculate the estimators $\hat{\alpha}^{*(s)}_{k}$ and $\hat{\beta}^{*(s)}_{k}$, together with the statistic $T^{*(s)}_{k}$ in (\ref{Eq2.15}).\\
$\bullet$ {\it Step 4}: Repeat {\it Steps 1-3} for S times (e.g. $S=500$), and calculate the critical values $q^{*(k)}_{1-\delta/2}$, $q^{*(k)}_{\delta/2}$,  $q^{*(k)}_{1-\delta/2d}$ and $q^{*(k)}_{\delta/2d}$. Output the confidence intervals as given in (\ref{Eq2.7}) and (\ref{Eq2.8}).
\end{algorithm}

%%%%%%%%%%%%%
\section{Divide-and-Conquer Sobel Test}\label{secDC-4}
\setcounter{equation}{0}

In this section, our focus is on estimating the mediation effects $\alpha_k\beta_k$'s and conducting multiple tests for
 \begin{eqnarray}\label{MT-36}
 H_{0k}: \alpha_k\beta_k=0~\leftrightarrow~ H_{Ak}: \alpha_k\beta_k\neq0,~~k=1,\cdots,d,
 \end{eqnarray}
within the context of big data. Assume that we have a set of independent and identically distributed (i.i.d.) samples, denoted as $\mathcal{D}_n=  \{(X_i,\mathbf{M}_i,Y_i,\mathbf{Z}_i)\}_{ i=1}^n$. One of the most commonly used methods for (\ref{MT-36}) is Sobel's test \cite[]{1982Asymptotic}, which employs
\begin{align}\label{Eq3.2}
T_k=\frac{\hat\alpha_k\hat\beta_k}{\hat\sigma_{\alpha_k\beta_k}}.
\end{align}
Here, the estimator $\hat\alpha_{k}$ is obtained through ordinary least squares regression with a variance of $\hat{\sigma}^2_{\alpha_k}$. The estimator $\hat\beta_{k}$ is  obtained through ordinary least squares regression for continuous outcomes and maximum likelihood estimation for binary outcomes, with a variance of $\hat{\sigma}^2_{\beta_k}$. According to \cite{1982Asymptotic}, the standard variance of $\hat\alpha_{k}\hat\beta_{k}$ is given by
\begin{align}\label{Sobel-43}
 \hat\sigma_{\alpha_k\beta_k}= \{\hat\alpha_k^2\hat\sigma^2_{\beta_k}+\hat\beta_k^2\hat\sigma^2_{\alpha_k}\}^{1/2},~ k=1,\cdots,d.
\end{align}
Let $\Omega=\{k: \alpha_k\beta_k \not=0, k=1, \cdots, d\}$ denote the index set of significant mediators. The  Bonferroni adjusted $p$-value for testing $H_{0k}:  \alpha_k\beta_k =0$  is calculated as
\begin{align}\label{Eq3.3}
P_k=2d\{1-\Phi(\vert T_k\vert )\},
\end{align}
where $T_k$ is  defined in (\ref{Eq3.2}) and $\Phi(\cdot)$ denotes  the cumulative distribution function of $N(0,1)$. The estimated index set of significant mediators is denoted as $\hat{\Omega}=\{k: P_k<0.05, k=1,\cdots,d\}$ under the significance level of 0.05.

When the sample size $n$ is extremely large, the computational burden of $T_k$ given in  (\ref{Eq3.2}) becomes onerous. {To  address this issue, we employ a divide-and-conquer approach (\citeauthor{Battey2018}, \citeyear{Battey2018}; \citeauthor{Shi2018}, \citeyear{Shi2018}; \citeauthor{Volgushev2019}, \citeyear{Volgushev2019})} when computing  the statistics $T_k$'s. We partition the entire dataset $\mathcal{D}_n$ into $J$ disjoint subsets, denoted as $\mathcal{D}_n^{(1)},\cdots, \mathcal{D}_n^{(J)}$, and estimate parameters based on each subset $\mathcal{D}_n^{(j)}$ randomly selected from the full data. For $j=1,\cdots,J$, let $\hat{\alpha}^{(j)}_{k}$ denote the least squares estimator obtained from $\mathcal{D}_n^{(j)}$, and let $\hat{\beta}^{(j)}_{k}$ denote the maximum likelihood estimator obtained from $\mathcal{D}_n^{(j)}$. Furthermore, the estimated standard errors of $\hat{\alpha}^{(j)}_{k}$ and $\hat{\beta}^{(j)}_{k}$ are denoted by $ \hat{\sigma}^{(j)}_{\alpha_{k}}$ and $ \hat{\sigma}^{(j)}_{\beta_{k}}$, respectively.
Similar to (\ref{Sobel-43}), we obtain $\hat{\sigma}^{(j)}_{\hat\alpha_{k}\hat\beta_{k}}=[\{\hat\alpha^{(j)}_{k}\hat\sigma^{(j)}_{\beta_{k}}\}^{2}+\{\hat\beta^{(j)}_{k}\hat\sigma^{(j)}_{\alpha_{k}}\}^{2}]^{1/2}$ for $k=1,\cdots,d$ and $j=1,\cdots,J$.  The final estimator for $M_k$'s mediation effect is
\begin{align}\label{Eq3.4}
\hat{\alpha}_{k}\hat{\beta}_{k}=\frac{1}{J}\sum_{j=1}^J\hat{\alpha}^{(j)}_{k}\hat{\beta}^{(j)}_{k},
\end{align}
and its estimated standard error is given by
\begin{align}\label{Eq3.5}
\hat{\sigma}_{\alpha_k\beta_k}=\frac{1}{J}\Big\{\sum_{j=1}^J(\hat{\sigma}^{(j)}_{\hat\alpha_{k}\hat\beta_{k}})^2\Big\}^{1/2}.
\end{align}
Algorithm \ref{alg2} presents a summary of the proposed divide-and-conquer Sobel test for mediation effects.
\begin{algorithm}[H]
\caption{ Divide-and-Conquer Sobel Test }\label{alg2}
$\bullet$ {\it Step 1}: Divide the entire dataset $\mathcal{D}_n$  into $J$ disjoint subsamples  of equal size  $n/J$, denoted as $\mathcal{D}_n^{(1)}$, $\mathcal{D}_n^{(2)}$, $\cdots$ and $\mathcal{D}_n^{(J)}$, respectively.\\
$\bullet$ {\it Step 2}: For each $\mathcal{D}_n^{(j)}$, we calculate the estimators $\hat{\alpha}^{(j)} _{k}$, $\hat{\beta}^{(j)}_{k}$ and $ \hat{\sigma}^{(j)}_{\alpha_{k}\beta_{k}}$. The final estimators of $\hat{\alpha}_{k}\hat{\beta}_{k}$ and $ \hat{\sigma}_{\alpha_k\beta_k}$ are given in (\ref{Eq3.4}) and (\ref{Eq3.5}), respectively.\\
$\bullet$ {\it Step 3}: Conduct multiple tests for the null hypothesis  $H_{0k}:  \alpha_k\beta_k =0$, where $k=1,\cdots,d$. If $P_k$ is less than 0.05, reject the  hypothesis $H_{0k}$ based on (\ref{Eq3.3}). i.e., $\hat{\Omega}=\{k: P_k<0.05, k=1,\cdots,d\}$.
\end{algorithm}

%%%%%%%%%%

\section{Simulation Study }\label{sec-sim}
\setcounter{equation}{0}
\subsection{SDB Confidence Intervals}\label{secSDB-S1}
\setcounter{equation}{0}
In this section, we perform simulations to assess the performance of SDB-based confidence intervals as discussed in Section \ref{sec2}. The mediators $M_k$'s are generated from
\begin{align}\label{SM1}
M_k = c_k + \alpha_k X + {\boldsymbol\eta}_k^\prime \mathbf{Z} + e_k,~~~ k = 1, \cdots, d.
\end{align}
We consider two  types  of outcomes (continuous and binary) as follows:\\
$\bullet$ Linear model:
\begin{eqnarray}\label{SL1}
Y &=& c + \gamma X + \beta_1 M_1 + \cdots + \beta_d M_d + \bm{\theta}^\prime \mathbf{Z} + \epsilon;\label{Eq2.1}
\end{eqnarray}
$\bullet$ Logistic model:
\begin{eqnarray}\label{LogM1}
P\{Y=1\} = \frac{\exp\{\gamma X + \beta_1 M_1 + \cdots + \beta_d M_d + \bm{\theta}^\prime \mathbf{Z}\}}{1+\exp\{\gamma X + \beta_1 M_1 + \cdots + \beta_d M_d + \bm{\theta}^\prime \mathbf{Z}\}}.
\end{eqnarray}

With the help of  R software, we generate random samples from models (\ref{SM1}), (\ref{SL1}) and (\ref{LogM1}), where {\black $\bm\alpha=(0, 0.2, 0, 0.1, 0.15)^\prime$, $\bm\beta=(0, 0, 0.2, 0.1, 0.15)^\prime$,} $c=c_k=\gamma=0.5$ and $\bm{\theta}= {\boldsymbol\eta}_k =(1,1)^\prime$. We generate $\mathbf{e}=(e_1,e_2,e_3,e_4,e_5)^\prime$ from $N_{5}(0,\bm\Sigma)$, where $\bm\Sigma=(0.5^{|i-j|})_{}$.  $\epsilon$ follows from $N(0,4)$, $\mathbf{Z}=(Z_1,Z_2)^\prime$ with $ Z_1$ and $ Z_2$ being generated from $N(0,2)$. We consider three different scenarios for generating exposure X:

Case 1: The variable $X$ is normally distributed with mean 0 and standard deviation 1.

Case 2: The variable $X$ is derived from the t-distribution with 5 degrees of freedom.

Case 3: The variable $X$ is generated from an exponential distribution with rate 1. \\
{\black As suggested by the reviewer, it would be informative to compare the performance of SDB confidence intervals with that of conventional percentile Bootstrap-based confidence intervals using full data (denoted as ``Bootstrap"). The subset size for the SDB method is chosen as $b=n^{0.7}$, where the full data sample size is $n=10^5$. The results presented in Tables~\ref{tab-1}-\ref{tab-4} are based on 500 repetitions.

In Table~\ref{tab-1}, we present the coverage probabilities (CPs) of individual mediators' 95\% confidence intervals obtained through SDB and Bootstrap methods, respectively.
Based on the findings presented in Table~\ref{tab-1}, all CPs are acceptable compared to 0.95, except for scenarios in which both the exposure-mediator and mediator-outcome effects are zero.  Ideally,  the coverage probability is 0.95, where a coverage probability between 0.925 and 0.975 is deemed acceptable based on Bradley's liberal robustness criterion(\citeauthor{Bradley-1978}, \citeyear{Bradley-1978}; \citeauthor{Valente-SEM-2017}, \citeyear{Valente-SEM-2017}; \citeauthor{Milica-MBR-2021}, \citeyear{Milica-MBR-2021}). Based on this criterion, the SDB and Bootstrap perform similar in terms of CPs.

 In Table~\ref{tab-2}, we present the mean lengths of estimated confidence intervals for the SDB and Bootstrap methods. It can be observed from Table~\ref{tab-2} that the SDB method yields slightly longer confidence intervals than the Bootstrap method. However, these differences are very small and negligible in practice. To investigate the performance  of Bonferroni-adjusted confidence intervals for both SDB and Bootstrap methods, we present the CPs and average lengths of estimated confidence intervals in Tables \ref{tab-3} and \ref{tab-4}, which yield comparable conclusions to those drawn from Tables \ref{tab-1} and \ref{tab-2}. In terms of precision, the proposed SDB method is comparable to the Bootstrap method in constructing confidence intervals for mediation effects.
}

\begin{table}[htp]
\begin{center}
\caption {The coverage probabilities of estimated confidence intervals with level 0.95$^\dagger$.}\label{tab-1} % Table 1
 %{{\bf Table 1}. The coverage probabilities of confidence intervals.} \\
\vspace{0.1in}
\normalsize
\begin{tabular}{clcccccc}
\hline
 & && SDB &&&Bootstrap&\\
\cmidrule(lr){3-5} \cmidrule(lr){6-8}
 Models& $(\alpha_k,\beta_k)$  & Case I & Case II  & Case III& Case I & Case II  & Case III  \\
\hline
Linear  &$(\alpha_1,\beta_1)=(0,0)$    &1    &1      & 1      &0.996 &1      & 0.992\\
         &$(\alpha_2,\beta_2)=(0.2,0)$ &0.956&0.944  & 0.972  &0.952 &0.942  & 0.962\\
        &$(\alpha_3,\beta_3)=(0,0.2)$  &0.956&0.948  & 0.942  &0.948 &0.934  & 0.932\\
        &$(\alpha_4,\beta_4)=(0.1,0.1)$&0.924&0.948  & 0.964  &0.922 &0.938  & 0.958\\
        &$(\alpha_5,\beta_5)=(0.15,0.15)$&0.958&0.958& 0.950  &0.950 &0.956  & 0.944\\
\hline
Logistic &$(\alpha_1,\beta_1)=(0,0)$   &1    &1      & 1      &1     &1      & 1\\
         &$(\alpha_2,\beta_2)=(0.2,0)$ &0.956&0.946  & 0.964  &0.964 &0.938  &  0.960\\
        &$(\alpha_3,\beta_3)=(0,0.2)$  &0.972&0.960  & 0.972  &0.952 &0.944  & 0.938\\
       &$(\alpha_4,\beta_4)=(0.1,0.1)$ &0.956&0.946  & 0.956  &0.944 &0.936  & 0.954\\
       &$(\alpha_5,\beta_5)=(0.15,0.15)$&0.962&0.968 & 0.948  &0.948 &0.964  & 0.944\\
\hline
\end{tabular}
\end{center}
{\vspace{-0.2cm}  \hspace{0cm}\footnotesize $\dagger$ ``SDB" denotes our method with subsampled double bootstrap algorithm; ``Bootstrap" denotes the conventional  percentile Bootstrap method with full data.
}\\
\end{table}

%---------------
\begin{table}[htp]
\begin{center}
\caption {The average lengths ($\times 10^{4}$) of estimated confidence intervals with level 0.95$^\dagger$.} \label{tab-2}% Table 2
 %{{\bf Table 1}. The coverage probabilities of confidence intervals.} \\
\vspace{0.1in}
%\small
\begin{tabular}{llccccccc}
\hline
 & && SDB &&&Bootstrap&\\
\cmidrule(lr){3-5} \cmidrule(lr){6-8}
 Models& $(\alpha_k,\beta_k)$  & Case I & Case II  & Case III& Case I & Case II  & Case III  \\
\hline
Linear  &$(\alpha_1,\beta_1)=(0,0)$    &6.698&5.196  &6.695  &1.362 &1.075  &1.399\\
         &$(\alpha_2,\beta_2)=(0.2,0)$ &49.71&49.43  &49.75  &49.22 &49.10  &49.26\\
        &$(\alpha_3,\beta_3)=(0,0.2)$  &25.58&19.89  &25.60  &24.59 &19.07  &24.55\\
        &$(\alpha_4,\beta_4)=(0.1,0.1)$&28.47&26.99  &28.58  &27.55 &26.36  &27.60\\
        &$(\alpha_5,\beta_5)=(0.15,0.15)$&41.77&39.81&41.78  &41.41 &39.53  &41.19\\
\hline
Logistic &$(\alpha_1,\beta_1)=(0,0)$   &9.972&7.764  & 10.07  &2.041 &1.547  & 2.054\\
         &$(\alpha_2,\beta_2)=(0.2,0)$ &73.57&73.45  & 74.34  &72.22 &72.62  & 72.95\\
        &$(\alpha_3,\beta_3)=(0,0.2)$  &27.05&20.93  & 26.93  &24.73 &19.09  & 24.49\\
       &$(\alpha_4,\beta_4)=(0.1,0.1)$ &39.95&38.63  & 40.41  &38.29 &37.64  & 38.73\\
&$(\alpha_5,\beta_5)=(0.15,0.15)$      &58.95&57.42  & 59.40  &57.63 &56.69  & 58.02\\
\hline
\end{tabular}
\end{center}
{\vspace{-0.2cm}  \hspace{0cm}\footnotesize $\dagger$ The meanings of ``SDB" and ``Bootstrap" are given in Table \ref{tab-1}. The numbers are increased by $10^4$ times.
}\\
\end{table}

%%%%%%%%%%%%%%%%%%%%%%%%%%%
\begin{table}[htp]
\begin{center}
\caption {The coverage probabilities of Bonferroni adjusted confidence intervals$^\dagger$.} \label{tab-3}
 %{{\bf Table 2}. The coverage probabilities of joint confidence intervals.} \\
\vspace{0.1in}
\normalsize
\begin{tabular}{lcccccccccc}
\hline
  &&SDB &&& Bootstrap&\\
\cmidrule(lr){2-4} \cmidrule(lr){5-7}
 Models& Case I& Case II & Case III&Case I & Case II & Case III \\
\hline
Linear    & 0.950 & 0.948 & 0.966 & 0.946 & 0.938 & 0.940\\
Logistic  & 0.970 & 0.964 & 0.958 & 0.956 & 0.934 & 0.952\\
\hline
\end{tabular}
\end{center}
{\vspace{-0.2cm}  \hspace{1.2cm}\footnotesize $\dagger$ The meanings of ``SDB" and ``Bootstrap" are given in Table \ref{tab-1}; The confidence level is 0.95.
}\\
\end{table}

%---------------
\begin{table}[htp]
\begin{center}
\caption {The average lengths ($\times 10^{4}$) of Bonferroni adjusted confidence intervals  with level 0.95$^\dagger$.} \label{tab-4}% Table 1
 %{{\bf Table 1}. The coverage probabilities of confidence intervals.} \\
\vspace{0.1in}
\normalsize
\begin{tabular}{clcccccc}
\hline
 & && SDB &&&Bootstrap&\\
\cmidrule(lr){3-5} \cmidrule(lr){6-8}
 Models& $(\alpha_k,\beta_k)$  & Case I & Case II  & Case III& Case I & Case II  & Case III  \\
\hline
Linear  &$(\alpha_1,\beta_1)=(0,0)$    &10.11&7.829  & 10.05  &2.047 &1.606  & 2.088\\
         &$(\alpha_2,\beta_2)=(0.2,0)$ &64.93&64.27  & 64.73  &63.83 &63.36  & 63.76\\
        &$(\alpha_3,\beta_3)=(0,0.2)$  &34.13&26.57  & 34.11  &31.73 &24.67  & 31.68\\
        &$(\alpha_4,\beta_4)=(0.1,0.1)$&37.86&35.38  & 37.97  &35.55 &33.92  & 35.59\\
        &$(\alpha_5,\beta_5)=(0.15,0.15)$&54.83&51.99& 54.53  &53.56 &50.99  & 53.01\\
\hline
Logistic &$(\alpha_1,\beta_1)=(0,0)$   &14.99&11.66  & 15.11  &3.058 &2.333  & 3.071\\
         &$(\alpha_2,\beta_2)=(0.2,0)$ &95.77&95.48  & 97.31  &93.00 &93.91  & 94.47\\
        &$(\alpha_3,\beta_3)=(0,0.2)$  &36.93&28.58  & 36.89  &31.93 &24.71  & 31.64\\
       &$(\alpha_4,\beta_4)=(0.1,0.1)$ &53.25&51.14  & 53.90  &49.56 &48.57  & 50.05\\
&$(\alpha_5,\beta_5)=(0.15,0.15)$      &77.33&75.09  & 77.82  &74.23 &73.34  & 74.99\\
\hline
\end{tabular}
\end{center}
{\vspace{-0.2cm}  \hspace{0cm}\footnotesize $\dagger$ The meanings of ``SDB" and ``Bootstrap" are given in Table \ref{tab-1}. The numbers are increased by $10^4$ times.
}\\
\end{table}

We perform a second simulation to compare the computational efficiency of the proposed SDB method with that of the traditional  percentile Bootstrap method.
The data is generated as the first simulation with Case I, except that $\bm\alpha=(0.5,\cdots,0.5)^\prime$ and $\bm\beta=(0.5,\cdots,0.5)^\prime$ for $d=5$, 10 and 20, respectively. The subset size for SDB method is determined as  $b=n^{0.7}$, where the full data size is $n=10^5$. The computation is implemented by a computer with 64GB memory (running R code). In Table~\ref{tab-5}, we present the average computation times (in seconds) for SDB and Bootstrap methods based on 10 repetitions, without taking into account data generation.  The results demonstrate that the proposed SDB method exhibits significantly higher computational efficiency compared to the Bootstrap procedure.
By \cite{Sengupta2016A},  increasing the size of the subset $b$ results in greater benefits in terms of statistical accuracy, but at a higher computational cost.  For a given computational time budget,  it is still  unclear how to choose an optimal subset size that balances statistical accuracy and running time for SDB procedure.  From a practical perspective, the practitioner may choose the largest possible subset size within a given computational time budget.

\begin{table}[htp]
\begin{center}
\caption {The comparison of computation time (in seconds) with Case I$^\dagger$.}\label{tab-5}
%{{\bf Table 3}. The comparison of computation time with Case I.} \\
\vspace{0.1in}
\normalsize
\begin{tabular}{lcccccccccc}
\hline
&& SDB&&&&Bootstrap\\
\cline{2-4} \cline{6-8}
Models&  $d=5$  &  $d=10$ &  $d=20$ &&$d=5$  &  $d=10$ &  $d=20$ \\
\hline
 Linear      & 16.292 &26.615 & 48.951 && 180.72 &326.38  & 633.98\\
Logistic     & 20.211 &32.108 & 60.498 && 299.17 &472.72  & 917.27\\
\hline
\end{tabular}
\end{center}
{\vspace{-0.2cm}  \hspace{1.5cm}\footnotesize $\dagger$ The meanings of ``SDB" and ``Bootstrap" are given in Table \ref{tab-1}; { $d$ is the number of mediators}.}
\end{table}

\subsection{DC-based Sobel Test}

In this section, we perform simulations to assess the efficacy of the Sobel test method based on divide-and-conquer (DC) approach. {\black  We consider two types of outcomes and generate random samples from models (\ref{SM1}), (\ref{SL1}) and (\ref{LogM1}). The settings are similar to those in Section \ref{secSDB-S1}, except that a linear model is used with
$\bm\alpha=(0,0.1,0,0.025,0.035)^\prime$ and $\bm\beta=(0,0,0.15,0.025,0.035)^\prime$, as well as a  logistic model with $\bm\alpha=(0, 0.15, 0, 0.025, 0.035)^\prime$ and $\bm\beta=(0,0,0.15,0.035,0.05)^\prime$.
 The  sample size for full data is chosen as $n=10^5$.} The number of blocks are $J=1$, 5, 50 and 100, respectively. We would like to emphasize that the scenario with $J = 1$ is based on the entire sample, without any subsampling, and thus corresponds to a standard analysis. The results in Tables \ref{tab-6}-\ref{tab-9} are based on 500 repetitions.

%-------------------------
\begin{table}[htp] % Table 4
\begin{center}
\caption {Bias ($\times 10^{6}$) and MSE ($\times 10^{8}$; in parenthesis) for mediation effects with Case 1$^\dagger$.}\label{tab-6}
% {{\bf Table 4}. Bias and MSE (in parenthesis) for mediation effects with $n=10^6$.} \\
\vspace{0.1in}
\normalsize
\begin{tabular}{llcccc}
\hline
Models& $(\alpha_k,\beta_k)$&    $J=1$        & $J=5$&$J=50$      & $J=100$ \\
\hline
Linear&$(0,0)$   &-0.2101  &1.4898  &1.7893   &-2.8824\\
&&               (0.0534) &(0.2237)&(1.7845) &(3.6766)\\
&$(0.1,0)$       &-14.388  &-13.879  &-21.925   &-17.121\\
&&               (36.628) &(36.547)&(39.541) &(41.749)\\
&$(0,0.15)$      &-43.309  & -45.686  &-46.746   &-39.853\\
&&               (23.569) &(23.747)&(24.905) &(27.193)\\
&$(0.025,0.025)$  &13.504  &14.310  &7.8653   &7.6704\\
&&              (2.9042) &(3.0187)&(4.8931) &(6.1670)\\
&$(0.035,0.035)$ &6.1792  &2.5440  &-8.8570   &-9.4346\\
&&              (6.7853) &(6.8542)&(8.9651) &(10.821)\\
\hline
Logistic&$(0,0)$&0.3281  &-1.6809  & -13.570   &-16.801\\
&&              (0.0798) &(0.4096)&(3.7180) &(8.1516)\\
&$(0.15,0)$      &-86.649  & -84.675  & -87.789 &-105.19\\
&&               (172.21) &(172.38)&(183.22) &(190.29)\\
&$(0,0.15)$     &-46.762  &-44.841  & -38.833   &-28.515\\
&&              (22.282) &(22.873)&(27.155) &(30.603)\\
&$(0.025,0.035)$ &14.052  &14.863  &19.785   &28.341\\
&&              (5.9699) &(6.4280)&(11.044) &(14.347)\\
&$(0.035,0.05)$ &2.0318  &1.7772  &16.018   & 38.367\\
&&              (12.071) &(12.633)&(16.705) &(21.584)\\
\hline
\end{tabular}
\end{center}
{\vspace{-0.2cm}  \hspace{2.3cm}\footnotesize $\dagger$$J=1$ is the same as a standard analysis without subsamples. The Bias and

\hspace{2.3cm} MSE are are increased by $10^6$ and $10^8$ times, respectively.
}\\
\end{table}
%%%%%%%%%%%%%%%%%%%%%%
Tables \ref{tab-6}-\ref{tab-8}  present the estimated bias (Bias), which is calculated as the average of mediation effect estimates minus the true value, and mean square error (MSE) of mediation effect estimates given by
\begin{eqnarray*}
{\rm MSE}(\alpha_k\beta_k)= \frac{1}{500} \sum_{j=1}^{500} \left\{\hat{\alpha}^{(j)}_k\hat{\beta}^{(j)}_k - \alpha_k\beta_k\right\}^2,
\end{eqnarray*}
where $\hat{\alpha}^{(j)}_k$ and $\hat{\beta}^{(j)}_k$ are the corresponding estimates from the $j$th repetition, $j=1, \cdots, 500$. It can be seen from Tables  \ref{tab-6}-\ref{tab-8} that all estimators are unbiased towards the true values of  mediation effects. {\black As the number of blocks increases, the differences in Bias (or MSE) are very small and can be considered negligible in practical applications.}

{Table \ref{tab-9} reports the results of {family-wise error rate (FWER)} and Power when conducting multiple testing with the Algorithm \ref{alg2}, where
\begin{align*}
H_{0k}: \alpha_k\beta_k\neq 0 ~\leftrightarrow ~H_{1k}: \alpha_k\beta_k\neq 0,~k=1,\cdots,5,
\end{align*}
at a significance level of 0.05}. Specifically, the Power and FWER are defined as
\begin{align}\label{Eq4.5}
{\rm Power}= \frac{1}{|\Omega|}\sum_{k\in \Omega}\mathbb{P}(P_{k}<0.05),
\end{align}
and
\begin{align}\label{Eq4.6}
{\rm FWER } = \mathbb{P}(\exists k\in \Omega^c : P_{k}<0.05),
\end{align}
where $\Omega = \{4, 5\}$, $\Omega^c= \{1, 2, 3\}$ and $P_{k}$'s are obtained by the divide-and-conquer algorithm.  The results in Table \ref{tab-9} indicate that Power decreases as the number of blocks increases. The FWERs are much smaller than the significance level of 0.05.
{\black Essentially, our method involves a tradeoff between statistical efficiency and computational efficiency, which is a common issue for divide-and-conquer based procedures.}
%%%%%%%%%%%%%%
\begin{table}[htp] % Table 5
\begin{center}
\caption {Bias ($\times 10^{6}$) and MSE ($\times 10^{8}$; in parenthesis) for mediation effects with Case 2$^\dagger$.}\label{tab-7}
% {{\bf Table 4}. Bias and MSE (in parenthesis) for mediation effects with $n=10^6$.} \\
\vspace{0.1in}
\normalsize
\begin{tabular}{llcccc}
\hline
Models& $(\alpha_k,\beta_k)$&    $J=1$        & $J=5$&$J=50$      & $J=100$ \\
\hline
Linear&$(0,0)$   &0.6877  &1.2694  &0.1059   &5.2722\\
&&               (0.0241) &(0.1211)&(1.3666) &(2.5342)\\
&$(0.1,0)$       &-1.1405  &-1.988  &-0.3675   &-2.9085\\
&&               (41.227) &(41.337)&(42.253) &(44.342)\\
&$(0,0.15)$      &3.8327  &1.6769  &3.6794   &2.1754\\
&&               (13.338) &(13.310)&(15.247) &(16.092)\\
&$(0.025,0.025)$  &3.0014  & 5.3530  &5.4643   &7.1933\\
&&              (3.0211) &(3.0332)&(4.0692) &(5.4192)\\
&$(0.035,0.035)$ &-8.7985  &-8.7657  &-5.1591   & -4.1719\\
&&              (6.1348) &(6.2202)&(7.6394) &(8.8628)\\
\hline
Logistic&$(0,0)$&-1.0418  &-3.3285  &-5.7510   &-7.6715\\
&&              (0.0417) &(0.2051)&(2.4418) &(4.6024)\\
&$(0.15,0)$      &19.296  & 19.469  &17.297   &10.674\\
&&               (168.19) &(168.59)&(174.62) &(181.89)\\
&$(0,0.15)$     &8.8453  & 10.717  &11.768   &4.7968\\
&&              (12.992) &(13.215)&(16.153) &(17.771)\\
&$(0.025,0.035)$  &2.8607  & 2.8227  &15.713   &23.402\\
&&              (5.4237) &(5.6156)&(7.7186) &(10.395)\\
&$(0.035,0.05)$ &-2.6734  & -1.8057  &9.0643   &11.053\\
&&              (10.983) &(11.058)&(14.024) &(16.714)\\
\hline
\end{tabular}
\end{center}
{\vspace{-0.2cm}  \hspace{2.3cm}\footnotesize $\dagger$$J=1$ is the same as a standard analysis without subsamples. The Bias and

\hspace{2.3cm} MSE are are increased by $10^6$ and $10^8$ times, respectively.
}\\
\end{table}

%%%%%%%%%%%%%%
\begin{table}[htp] % Table 6
\begin{center}
\caption {Bias ($\times 10^{6}$) and MSE ($\times 10^{8}$; in parenthesis) for mediation effects with Case 3$^\dagger$.}\label{tab-8}
% {{\bf Table 4}. Bias and MSE (in parenthesis) for mediation effects with $n=10^6$.} \\
\vspace{0.1in}
\normalsize
\begin{tabular}{llcccc}
\hline
Models& $(\alpha_k,\beta_k)$&    $J=1$        & $J=5$&$J=50$      & $J=100$ \\
\hline
Linear&$(0,0)$   &-0.5258  &-0.7788  &-1.3620 & -4.4085\\
&&               (0.0359) &(0.2017)&(1.8676) &(3.7627)\\
&$(0.1,0)$       &2.4277  &1.4731  & -9.7915   & -12.549\\
&&               (41.698) &(42.180)&(44.403) &(45.680)\\
&$(0,0.15)$      &5.5580  &6.9381  &9.8424   &1.7548\\
&&               (19.516) &(19.412)&(21.151) &(24.043)\\
&$(0.025,0.025)$  &-7.7499  &-7.3152  &-16.176   &-12.278\\
&&              (3.2775) &(3.5264)&(5.5125) &(8.0929)\\
&$(0.035,0.035)$ &-2.8638  &-6.5763  & -5.0265   & -0.2281\\
&&              (5.4613) &(5.5391)&(7.4047) &(10.017)\\
\hline
Logistic&$(0,0)$&1.7192  & 2.8207  &0.4644   &-10.200\\
&&              (0.0691) &(0.4125)&(4.1914) &(8.6075)\\
&$(0.15,0)$      &-34.695  &-34.809  &-27.133   &-11.657\\
&&               (173.65) &(172.98)&(181.44) &(191.11)\\
&$(0,0.15)$     &-41.219  & -42.612  &-53.293   &-67.098\\
&&              (23.108) &(23.644)&(28.319) &(32.728)\\
&$(0.025,0.035)$  &-12.907  &-12.109  &-7.6015   &-16.501\\
&&              (5.9377) &(6.1271)&(9.8584) &(14.336)\\
&$(0.035,0.05)$ &15.605  &21.180 &36.543   &51.505\\
&&              (11.948) &(12.029)&(15.972) &(21.353)\\
\hline
\end{tabular}
\end{center}
{\vspace{-0.2cm}  \hspace{2.3cm}\footnotesize $\dagger$$J=1$ is the same as a standard analysis without subsamples. The Bias and

\hspace{2.3cm} MSE are are increased by $10^6$ and $10^8$ times, respectively.
}\\
\end{table}

\begin{table}[htp] % Table 7
\begin{center}
\caption {{The family-wise error rate (FWER)} and Power for multiple testing$^\dagger$.}\label{tab-9}
% {{\bf Table 5}. FWER and Power for the mediation effect with $n=10^6$  } \\
\vspace{0.1in}
\normalsize
\begin{tabular}{lcccccccccc}
\hline
Models&Settings&&     $J=1$       & $J=5$  & $J=50$      &$J=100$  \\
\hline
Linear &Case 1&FWER   & 0.024&0.020&0.018&0.010\\
              &&Power & 0.954&0.935&0.649&0.476\\
       &Case 2&FWER   & 0.016&0.016&0.014&0.010\\
              &&Power & 0.945&0.936&0.770&0.635\\
       &Case 3&FWER   & 0.018& 0.018&0.018&0.018\\
              &&Power & 0.937&0.914&0.658& 0.480\\
\hline
Logistic&Case 1&FWER  & 0.028&0.028&0.024&0.022\\
              &&Power & 0.952&0.924&0.667&0.511\\
       &Case 2&FWER   & 0.020& 0.020&0.010&0.008\\
              &&Power &0.955 &0.946&0.787&0.636\\
       &Case 3&FWER   &0.014&0.018& 0.020&0.016\\
              &&Power & 0.934&0.915&0.655&0.500\\
\hline
\end{tabular}
\end{center}
{\vspace{-0.2cm}  \hspace{2.3cm}\footnotesize $\dagger$$J=1$ is the same as a standard analysis without subsamples.
}\\
\end{table}

We conduct the second simulation to investigate the computational efficiency of Algorithm \ref{alg2}.  The generation of data is the same as the first simulation with Case 1, except that $\bm\alpha=(0.5,\cdots,0.5)^\prime$ and $\bm\beta=(0.5,\cdots,0.5)^\prime$, where $d=5$, 50 and 100, respectively. The computation is carried out by a computer with 64GB of memory.  To simulate scenarios involving multiple machines simultaneously, we vary
the sample sizes as $n$, $n/5$, $n/50$ and $n/100$ corresponding to $J$ values of 1, 5, 50 and 100, respectively. Here $J=1$ is actually the conventional analysis of all the data (i.e., no subsamples). In Table \ref{tab-10}, we report the computation times of  Algorithm \ref{alg2} without accounting for data generation.  These times are the average of 10 repetitions and are measured in seconds. According to Table~\ref{tab-10}, the computational efficiency of the divide-and-conquer algorithm increases with the utilization of additional computational resources. In consideration of the tradeoff between statistical efficiency and computational speed, it is recommended to minimize the use of computational resources within the specified time budget.

%%%%%%%%%%%%%%%%%%%%%%%%%%%
\begin{table}[htp] % Table 8
\begin{center}
\caption{The comparison of computation time (in seconds) with Case 1$^\dagger$.}\label{tab-10}
%{{\bf Table 6}. The comparison of computation time with $n=10^6$. } \\
\vspace{0.1in}
\normalsize
\begin{tabular}{llcccccccccc}
\hline
Models  &dimension&   $J=1$   & $J=5$   &   $J=50$   & $J=100$ \\
\hline
Linear    &$d=5$   &0.352 & 0.071 &0.016 &0.014 \\
          &$d=50$  &3.292 & 0.575 &0.092 &0.070\\
          &$d=100$ &6.543 & 1.116 &0.208 &0.149\\
\hline
Logistic  &$d=5$   &0.785 & 0.155 &0.025 &0.015\\
          &$d=50$  &7.044 & 1.389 &0.238 &0.133\\
          &$d=100$ &21.279 &3.735 &0.607 &0.335\\
\hline
\end{tabular}
\end{center}
{\vspace{-0.2cm}  \hspace{1.5cm}\footnotesize $\dagger$$J=1$ is the same as a standard analysis without subsamples; d is the number of mediators.}\\
\end{table}
%%%%%%%%%%%%%%%%
\section{Application}\label{sec-data}
\setcounter{equation}{0}
\subsection{Continuous Outcomes }\label{sec5.1}
In this section, we apply the  SDB-based confidence
intervals and DC-based Sobel Test  to a large dataset about the P2P lending platform. The P2P is the abbreviation of peer to peer, which is a new type of network lending platform. The dataset is about the loan transaction data of Lending Club Company from 2007 to 2015, which is public available at  https://www.kaggle.com/wendykan/lending-club-data. {\cite{herzenstein2008-SMR} demonstrated a significant correlation between the borrower's housing property and interest rates. It can be reasonably anticipated that the total loan amount  and repayment period are correlated with borrower's housing property,  while the interest rate is correlated with both loan amount and repayment period. Therefore, it is interesting to investigate whether loan amount and repayment period are along the causal pathways from the borrower's housing property to the interest rate.}  After eliminating missing values from the dataset, the resulting sample size is $n=842,322$.

For the purpose of analysis, the exposure variable $X$ represents the borrower's housing property, which includes rental properties, owned properties and mortgaged properties. Specifically, we define $X = -1$ for rented or mortgaged properties and $X = 1$ for owned properties. The variable $Y$ represents the interest rate on the loan. It is expected that the status ``rent" or ``mortgage" should has an opposite effect on $Y$ compared to the status ``own". To characterize this attribute, we represent the binary exposure $X$ as $-1$ and 1 rather than encoding it as 0 and 1. The mediator $M_1$ represents the total amount of the loan (in  units of ten thousand dollars). The mediator  $M_2$ represents the repayment period of the loan, where $M_2=1$ if  the repayment period is 36 months and $M_2=2$ if the repayment period is 60 months.  Moreover, we designate $Z_1$ as the borrower's annual income (in units of ten thousand dollars), while $Z_2$ denotes the borrower's years of work experience.
Our purpose is to explore whether the two mediators, $M_1$ and $M_2$,  have significant mediated effects on the pathways linking  housing status (X)  to loan interest rates (Y).

We first employ the least squares method to fit the mediation linear models (as shown in Figure~\ref{fig-2}) with full data:
\begin{eqnarray*}
Y&=& 29.667- 0.025 X + 0.019 M_1 + 0.066 M_2 -0.009 Z_1-3.169 Z_2+ \epsilon,\label{Eq5.1}\\
M_1 &=& 1.745-0.009 X +0.043 Z_1-0.114 Z_2+e_1,\label{EqMM5.2}\\
M_2 &=& 2.082-0.009 X +0.006 Z_1-0.159 Z_2+e_2,\nonumber
\end{eqnarray*}

\begin{figure}[H]%Figure 2
\centering
\includegraphics[scale=0.6]{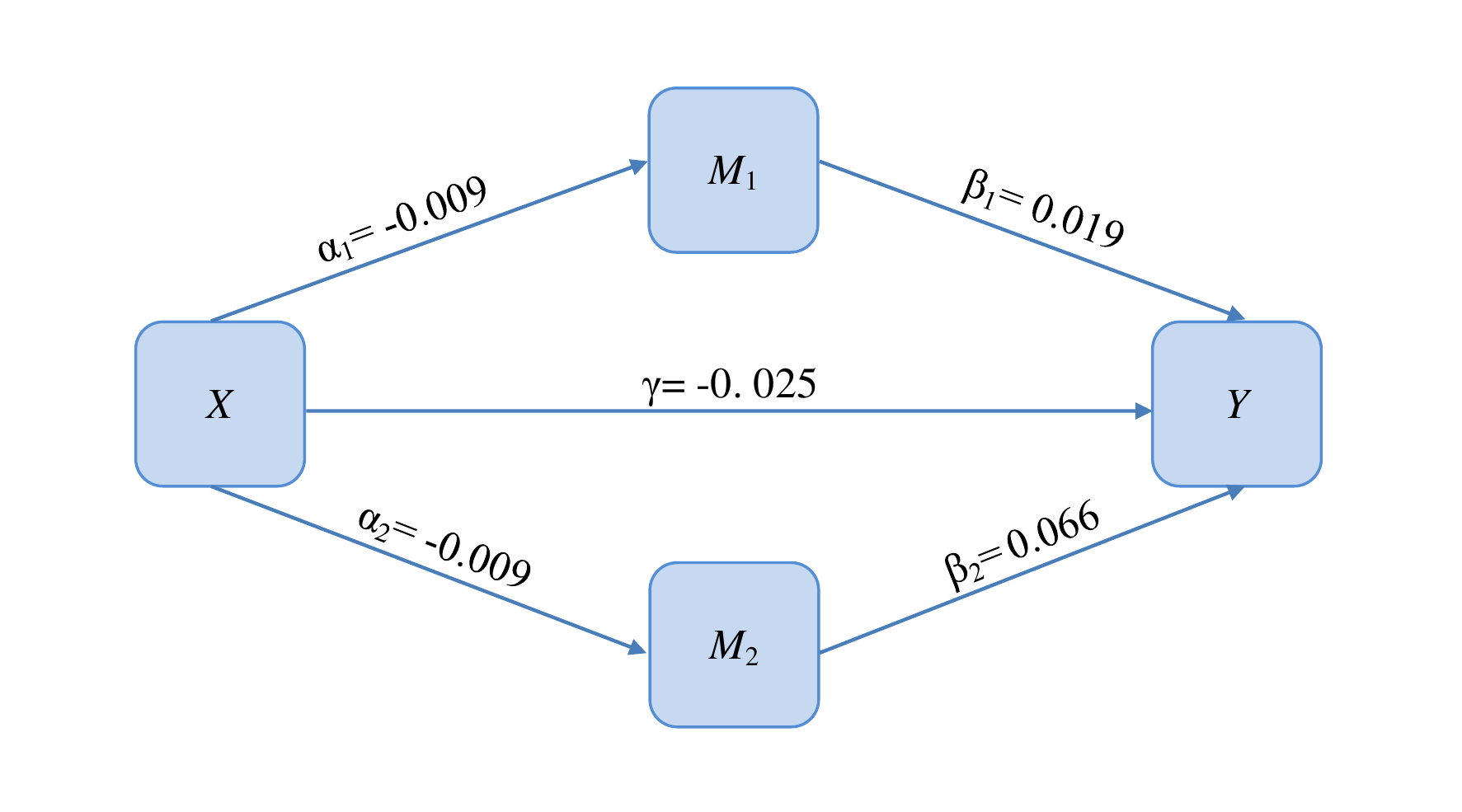}\\
\vspace{-1cm}
\begin{center}
\caption {The multiple mediators model in Section \ref{sec5.1} (omitted confounding variables).}\label{fig-2}
\end{center}
\end{figure}

%%%%%%%%%%%%%%%%%%%%%%%%%%%
\begin{table}[htp]
\begin{center}
\caption {The estimated confidence intervals of mediation effects in Section \ref{sec5.1}.}\label{tab-11}
 %{{\bf Table 7}. The SDB confidence intervals of mediation effects in section 5.1.} \\
\vspace{0.1in}
\normalsize
%\begin{threeparttable}
\begin{tabular}{llccccc}
\hline
Estimates & Methods& ${\rm CI_{single}}$& ${\rm CI_{adjusted}}$ \\
\hline
$\hat\alpha_1\hat\beta_1=-0.00017$& SDB & $[-0.00024, -0.00008]$&$[-0.00026, -0.00006]$ \\
& Bootstrap & $[-0.00024,  -0.00011]$&$[-0.00071,-0.00046]$\\
%& Sobel & $[-0.00023,-0.00010]$&$[-0.00024,-0.00010]$\\
\hline
$\hat\alpha_2\hat\beta_2=-0.00059$& SDB & $[-0.00071,-0.00046]$&$[-0.00073, -0.00043]$\\
& Bootstrap & $[-0.00071,  -0.00046]$&$[-0.00073, -0.00045]$\\
%& Sobel & $[-0.00071, -0.00047]$&$[-0.00073,-0.00045]$ \\
\hline
\end{tabular}
\end{center}
{\vspace{-0.2cm}  \hspace{0.1cm}\footnotesize $\dagger$ The meanings of ``SDB" and ``Bootstrap " are given in Table \ref{tab-1}.
}\\
\end{table}

According to Figure~\ref{fig-2}, after  adjusting for other variables, exposure $X$ has a negative direct effect ($\gamma = -0.025$) on the outcome $Y$. That is to say, when applying for a loan, homeowners would be eligible for a lower interest rate \cite[]{herzenstein2008-SMR}. Borrowers who possess a residential property exhibit a tendency towards requiring a smaller loan amount ($\alpha_1 = -0.009$) and opting for a shorter repayment period ($\alpha_2 = -0.009$). The interest rate has a positive correlation with borrower's loan amount ($\beta_1 = 0.019$) and  repayment period ($\beta_2= 0.066$), but it has a negative relationship with both annual incomes and working years.

%i.e., the borrower owning a house would be required less
%interest rate when applying for a loan \cite[]{herzenstein2008-SMR}. Those borrowers who own a house tend to need a smaller loan amount ($\alpha_1 = -0.009$) and a shorter  repayment period ($\alpha_2 = -0.009$). The interest rate has a positive correlation with borrower's loan amount ($\beta_1 = 0.019$) and  repayment period ($\beta_2= 0.066$). Moreover, the interest rate has a negative relationship with both annual incomes and working years.
\begin{table}[htp]
\begin{center}
\caption {Estimates, standard errors and $p$-values in Section \ref{sec5.1}$^\dagger$.}\label{tab-12}
%{{\bf Table 8}. Estimates, standard errors and P-values in Section \ref{sec5.2}.} \\
\vspace{0.1in}
\normalsize
\begin{tabular}{ccccccccccccccccc}
\hline
 &   $J=1$   & $J=3$   &   $J=5$   & $J=10$ \\
\hline
 $\hat\alpha_1\hat\beta_1$      &$-0.00017$ &-0.00106 &-0.00129 & -0.00047 \\
 $\hat{\sigma}_{\alpha_1\beta_1}$ &0.00003 &0.00013 &0.00018 & 0.00018 \\
  $P_{1}$                       &$< 10^{-5}$ &$< 10^{-5}$ &$< 10^{-5}$ & 0.01551\\
  \hline
 $\hat\alpha_2\hat\beta_2$   &$-$0.00059 & -0.00120 &-0.00131 &-0.00162 \\
 $\hat{\sigma}_{\alpha_2\beta_2}$&0.00006 &0.00010 & 0.00012 & 0.00017 \\
$P_{2}$ &$< 10^{-5}$ &$< 10^{-5}$ &$< 10^{-5}$ &$< 10^{-5}$\\
\hline
\end{tabular}
\end{center}
{\vspace{-0.2cm}  \hspace{3cm}\footnotesize $\dagger$  $J=1$ is the same as a standard analysis without subsamples.
}\\
\end{table}

We apply the proposed SDB approach to estimate confidence intervals for mediation effects, with a subset size of $b= n^{0.8}$ and a confidence level of 0.95. For comparison, we also provide the confidence intervals obtained by traditional percentile Bootstrap with full data. Table~\ref{tab-11} presents the estimated confidence intervals of mediation effects, indicating that both estimators for mediation effects significantly deviate from zero at a confidence level of 0.95. i.e., both the loan amount ($M_1$)  and  repayment period  ($M_2$) can be considered  as two significant mediators along the pathways from housing status $(X)$  to loan interest rates $(Y)$.

Furthermore, we employ the proposed divide and conquer method (the Algorithm \ref{alg2}) to estimate indirect effects and conduct multiple tests:
\begin{align*}
H_{0i}:~\alpha_i\beta_i = 0~\leftrightarrow~H_{1i}:~\alpha_i\beta_i \neq 0,~i=1~{\rm and}~2.
\end{align*}
The significance level is set at 0.05, while the number of blocks is selected as $J $ = 1, 3, 5 and 10 respectively.   Note that the scenario with $J=1$ is the same as a standard analysis without subsamples.  Table \ref{tab-12}  presents the estimates, standard errors, and p-values of indirect effects across various $J$ values.  As $J$ increases, the standard errors are observed to grow larger overall. This phenomenon provides evidence for the trade-off between statistical efficiency and computational speed inherent in divide-and-conquer algorithms.

\subsection{Binary Outcomes }\label{sec5.2}

In this section, we will employ our proposed two methods
for analyzing a large dataset about the loan data of Lending Club from 2007 to 2015,  which is  publicly available at  https://www.kaggle.com/wendykan/lending-club-data.
According to \cite{Kumar-2007-AMCIS}, the borrower's annual income is correlated with the total loan amount, while the loan status of the borrower is also associated with the total loan amount. In particular, borrowers with a higher loan amount are at an increased risk of defaulting. Additionally, it is anticipated that the borrower's annual income will have a correlation with the loan's interest rate, and those with higher rates may struggle to meet repayment deadlines (\citeauthor{serrano2015determinants}, \citeyear{serrano2015determinants}; \citeauthor{emekter2015evaluating}, \citeyear{emekter2015evaluating}).  A relevant question is whether loan amount and interest rate mediate the impact of annual income on default status. To achieve this objective, we define exposure $X$ as the borrower's annual income (in ten thousand dollars), and the binary outcome $Y$ represents the loan status of the borrower, where $Y = 1$ indicates default and $Y = 0$ indicates timely repayment. The mediator variables $M_1$ and $M_2$ represent the total loan amount (in ten thousand dollars) and the interest rate of the loan, respectively.  We set the repayment period as a covariate $Z_1$, where $Z_1=1$ for a period of 36 months  and $Z_1=2$ for a period of 60 months.  For the purpose of analysis, we have standardized the three continuous variables $X$, $M_1$ and $M_2$ with mean zero and unit variance, respectively.  After removing missed data, the full data sample size is $n=825,994$.

First we estimate the parameters in models (\ref{CLOG-1}) and (\ref{CLOG-2}) with full data, which are presented as follows (see Figure \ref{fig-3}):
\begin{eqnarray*}\label{Eq5.2}
P(Y_i=1) &=& \frac{\exp\{-4.260-0.367 X + 0.044 M_1 +2.225 M_2 - 0.323 Z_1\}}{1+\exp\{-4.260-0.367 X + 0.044 M_1 +2.225 M_2 - 0.323 Z_1\}},\\
M_1&=& 0.146+0.234 X +0.418 Z_1+e_1,\nonumber\\
M_2&=& 0.594-0.047 X +0.299 Z_1+e_2.\nonumber
\end{eqnarray*}

%-------------------------
\begin{figure}[H] %Figure 3
\centering
\includegraphics[scale=0.6]{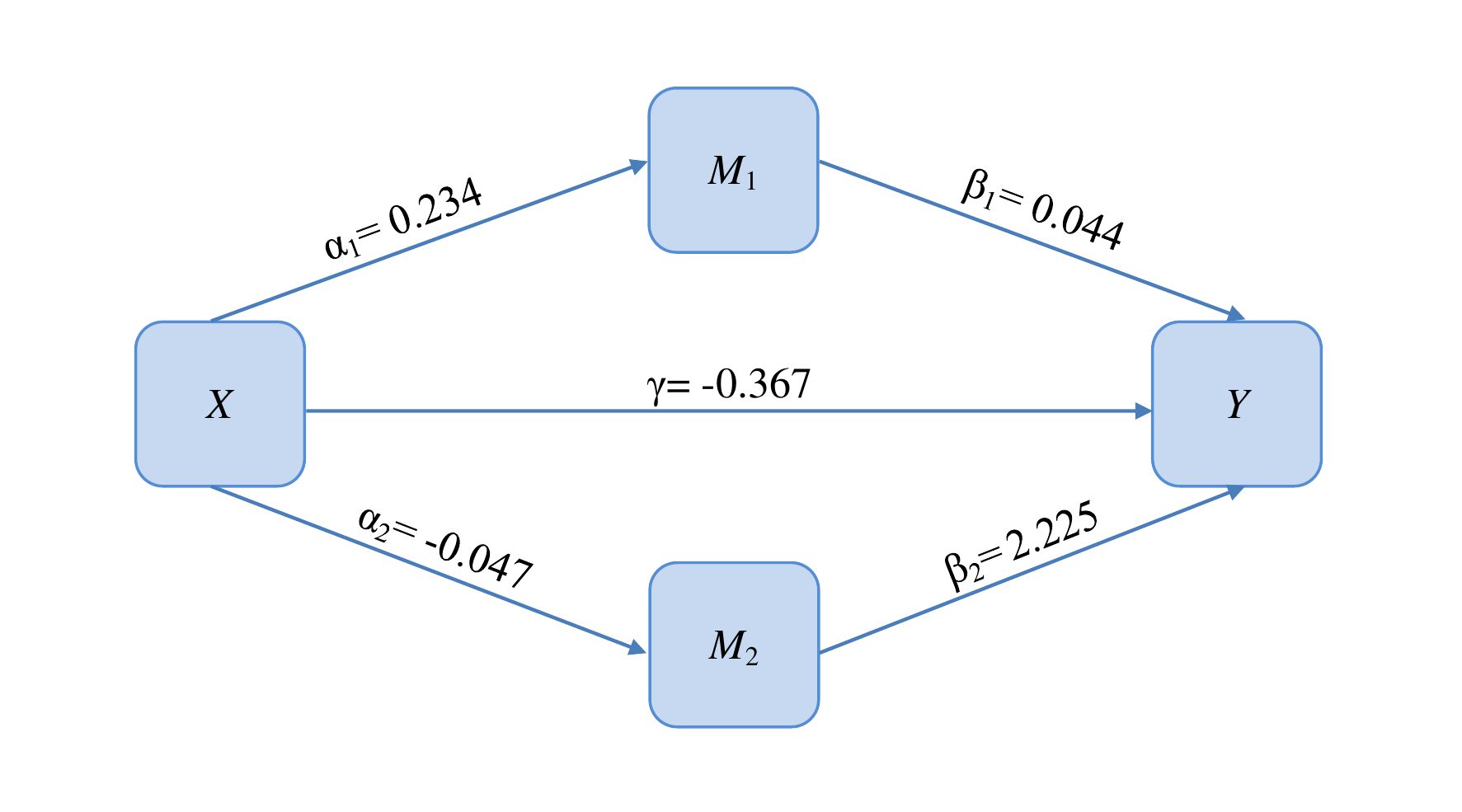}\\
\vspace{-1cm}
\begin{center}
\caption {The multiple mediators model in Section \ref{sec5.2} (omitted confounding variables).}\label{fig-3}
\end{center}
\end{figure}

{In view of Figure \ref{fig-3}, the exposure $X$ has a negative direct effect ($\gamma = -0.367$) on the outcome $Y$.  Although the interest rate may be lower for a borrower  with a high annual income ($\alpha_2 = -0.047$), an applicant with higher income possesses the confidence and capacity to  borrow larger amounts ($\alpha_1 = 0.234$). Essentially, the risk of default has a significant positive  correlation with the loan amount ($\beta_1 = 0.044$) and the interest rate ($\beta_2 = 2.225$). That is to say, if the loan amount and interest rate are higher, it may lead to borrowers defaulting on the loan.

}

\begin{table}[htp]
\begin{center}
\caption {The estimated confidence intervals of mediation effects in Section \ref{sec5.2}.}\label{tab-13}
 %{{\bf Table 7}. The SDB confidence intervals of mediation effects in section 5.1.} \\
\vspace{0.1in}
\normalsize
%\begin{threeparttable}
\begin{tabular}{llccccc}
\hline
Estimates & Methods& ${\rm CI_{single}}$& ${\rm CI_{adjusted}}$ \\
\hline
$\hat\alpha_1\hat\beta_1=0.01035$& SDB & $[0.00363, 0.01659]$&$[0.00204, 0.01753]$ \\
& Bootstrap & $[0.00392, 0.01753]$&$[0.00309,0.01853]$\\
%& Sobel & $[-0.00023,-0.00010]$&$[-0.00024,-0.00010]$\\
\hline
$\hat\alpha_2\hat\beta_2= -0.10453$& SDB & $[-0.11567,-0.09191]$&$[-0.11701, -0.08884]$\\
& Bootstrap & $[-0.12216, -0.09225]$&$[-0.12434,-0.09114]$\\
%& Sobel & $[-0.00071, -0.00047]$&$[-0.00073,-0.00045]$ \\
\hline
\end{tabular}
\end{center}
{\vspace{-0.2cm}  \hspace{0.8cm}\footnotesize $\dagger$ The meanings of ``SDB" and ``Bootstrap " are given in Table \ref{tab-1}.
}\\
\end{table}

In Table \ref{tab-13}, we implement the proposed SDB method to estimate confidence intervals for mediation effects. The subset size is set at $b = n^{0.8}$ and the confidence level is 0.95. In Table~\ref{tab-14}, we employ the proposed divide and conquer approach (Algorithm 2) to estimate indirect effects and conduct multiple tests:
\begin{align*}
H_{0i}:~\alpha_i\beta_i = 0~\leftrightarrow~H_{1i}:~\alpha_i\beta_i \neq 0,~i=1~{\rm and}~2.
\end{align*}
The significance level is set at 0.05,  and the number of blocks is  selected as $J=$ 1, 3, 5 and 10, respectively. Similar conclusions to those presented in Tables~\ref{tab-11} and \ref{tab-12} can be drawn from our findings here.

\begin{table}[htp]
\begin{center}
\caption {Estimates, standard errors and $p$-values in Section \ref{sec5.2}$^\dagger$.}\label{tab-14}
%{{\bf Table 8}. Estimates, standard errors and P-values in Section \ref{sec5.2}.} \\
\vspace{0.1in}
\normalsize
\begin{tabular}{ccccccccccccccccc}
\hline
 &   $J=1$   & $J=3$   &   $J=5$   & $J=10$\\
\hline
 $\hat\alpha_1\hat\beta_1$      &0.01035 & 0.01026 &0.01012 &0.01138 \\
 $\hat{\sigma}_{\alpha_1\beta_1}$ &0.00264 & 0.00275 &0.00273 &0.00288 \\
  $P_{1}$                       & 0.00018 &0.00038 &0.00043  &0.00016 \\
  \hline
 $\hat\alpha_2\hat\beta_2$    &-0.10453 &-0.10716&-0.10682 &-0.11106 \\
 $\hat{\sigma}_{\alpha_2\beta_2}$ &0.00128 &0.00131 &0.00131 &0.00135 \\
$P_{2}$           &$< 10^{-5}$ &$< 10^{-5}$ &$< 10^{-5}$&$< 10^{-5}$\\
\hline
\end{tabular}
\end{center}
{\vspace{-0.2cm}  \hspace{3.3cm}\footnotesize $\dagger$  $J=1$ is the same as a standard analysis without subsamples.
}\\
\end{table}

%%%%%%%%%%%%%%%%%%%%%
\section{Concluding Remarks}\label{sec-7r}

In this paper, we have investigated the statistical inference of mediation effects using large-scale datasets. First we proposed a SDB-based algorithm to construct confidence intervals of mediation effects. {Although we used $b=n^{0.7}$ for the SDB algorithm in the simulation, there are some other possible choices
for the subset sizes (e.g., $b=n^{0.6}$ or $b=n^{0.8}$) according to the available computing resources at hand.}  Then we presented a divide-and-conquer algorithm to estimate and test the mediation effects. Simulations and two  real-world examples were provided to demonstrate  the usefulness of our proposed methods in practical applications. {\black  As pointed out by the reviewer that we have considered the scenario of causally independent mediators for both SDB and divide-and-conquer algorithms. When the mediators are uncausally correlated, the product of individual parameters fails to fully capture individual causal indirect effects (\citeauthor{wang-SIM-2013}, \citeyear{wang-SIM-2013}; \citeauthor{2021-IJB-Allan}, \citeyear{2021-IJB-Allan}).
}

There exist several topics to be studied in the future. First,
the effect size measures  is an important topic for mediation analysis \cite[]{Preacher-2011-PM}, which is out of the scope for this manuscript.
It is interesting to consider the statistical inference for effect size measures in the context of massive data. Second,  the mediation analysis with exposure-mediator interaction is an important topic, such as \cite{Valeri-PM-2013} and \cite{Rijnhart-XM-2021}.  This manuscript is mainly focused on fast
calculation for mediation effects without interactions between exposure and mediators.
It is desirable to extend the proposed method to the situation that there exists
exposure-mediator interactions with massive data. Third, it is commonly required strong unconfoundedness assumptions for the identification of direct and indirect effects in mediation analysis. i.e., conventional mediation methods assume all confounders can be measured, which is often unverifiable in the case of large datasets. It is interesting to
investigate how to perform mediation analysis for massive dataset with hidden confounders \cite[]{WSDM-2022}. {\black Fourth, when the mediators are uncausally related, the joint indirect effect could be of main interest instead of the individual indirect effect (\citeauthor{wang-SIM-2013}, \citeyear{wang-SIM-2013}; \citeauthor{2021-IJB-Allan}, \citeyear{2021-IJB-Allan}). In this case, Sobel's multivariate delta method needs to take into account the covariance among the mediators \cite[]{1982Asymptotic}. Under the framework of massive data, it is interesting to consider how to perform valid
inference for the joint indirect effect with uncausally related mediators.

}

\section*{Acknowledgement}
The authors would like to thank the Editor, the Associate Editor and the reviewers for their constructive and insightful comments that greatly improved the manuscript.

\section*{Appendix}
\renewcommand{\theequation}{A.\arabic{equation}} % (S. equation)
\setcounter{equation}{0}
{
In this Appendix, we give the proof details of Theorem 2. Note that \cite{VanderWeele-AJE-2010} has provided the expressions of $\rm {NDE^{OR}}$, $\rm {NIE^{OR}}$ and $\rm {TE^{OR}}$ for logistic mediation model with one mediator. Taking the log scale on (\ref{ORNED}), we have
\begin{eqnarray}\label{A-1}
\log\left(\rm {NDE^{OR}}\right)&=&\log\left\{\frac{P\{Y(x,\mathbf{M}(x^*))=1\}/[1-P\{Y(x,\mathbf{M}(x^*))=1\}]}{P\{Y(x^*,\mathbf{M}(x^*))=1
\}/[1-P\{Y(x^*,\mathbf{M}(x^*))=1\}]}\right\}\\
&=& logit\left(P\{Y(x,\mathbf{M}(x^*))=1\}\right) - logit\left(P\{Y(x^*,\mathbf{M}(x^*))=1\}\right),\nonumber
\end{eqnarray}
where $logit(p)=\log(\frac{p}{1-p})$ for $p\in(0,1)$. Under the assumptions (C.1)-(C.4) and the outcome is rare, we get that
\begin{eqnarray}\label{A-2}
&&logit\left(P\{Y(x,\mathbf{M}(x^*))=1\}\right)\approx\log\left(P\{Y(x,\mathbf{M}(x^*))=1\}\right)\nonumber\\
&&= \log\left(\int_\mathbf{m}P\{Y(x,\mathbf{m})=1
|\mathbf{Z},\mathbf{M}(x^*)=\mathbf{m}\}P\{\mathbf{M}(x^*)=\mathbf{m}\}d\mathbf{m} \right)\nonumber\\
&&= \log\left(\int_\mathbf{m}P\{Y(x,\mathbf{m})=1
|\mathbf{Z}\}P\{\mathbf{M}(x^*)=\mathbf{m}|\mathbf{Z}\}d\mathbf{m} \right)\nonumber\\
&&= \log\left(\int_\mathbf{m}P\{Y=1
|x,\mathbf{m},\mathbf{Z}\}P\{\mathbf{M}=\mathbf{m}|x^*, \mathbf{Z}\}d\mathbf{m} \right)\nonumber\\
&&\approx \log\left(\int_\mathbf{m}\exp\{c+ \gamma x +\bbeta^\prime \mathbf{m} + \bm{\theta}^\prime \mathbf{Z}\}P\{\mathbf{M}=\mathbf{m}|x^*, \mathbf{Z}\}d\mathbf{m} \right)\nonumber\\
&&=\log\left(\exp\{c+ \gamma x  + \bm{\theta}^\prime \mathbf{Z}\}\int_\mathbf{m}\exp\{\bbeta^\prime \mathbf{m}\}P\{\mathbf{M}=\mathbf{m}|x^*, \mathbf{Z}\}d\mathbf{m} \right)\nonumber\\
&&=\log\left(\exp\{c+ \gamma x  + \bm{\theta}^\prime \mathbf{Z}\}E[e^{\bbeta^\prime \mathbf{M}}|x^*, \mathbf{Z}] \right)\nonumber\\
&&=\log\left(\exp\{c+ \gamma x  + \bm{\theta}^\prime \mathbf{Z}\}\exp\Big\{\bbeta^\prime \tilde{\mathbf{c}} + \balpha^\prime \bbeta x^* + \sum_{k=1}^d \beta_k{\boldsymbol\eta}_k^\prime \mathbf{Z}+\frac{1}{2}\bbeta^\prime {\boldsymbol\Sigma}_e \bbeta\Big\} \right)\nonumber\\
&& = c+ \gamma x  + \bm{\theta}^\prime \mathbf{Z} + \bbeta^\prime \tilde{\mathbf{c}} + \balpha^\prime \bbeta x^* + \sum_{k=1}^d \beta_k{\boldsymbol\eta}_k^\prime \mathbf{Z}+\frac{1}{2}\bbeta^\prime {\boldsymbol\Sigma}_e \bbeta,
\end{eqnarray}
where $\tilde{\mathbf{c}} = (c_1,\cdots,c_d)^\prime$, and ${\boldsymbol\Sigma}_e$ is the covariance matrix of mean-zero normal vector $\mathbf{e}=(e_1,\cdots,e_d)^\prime$ in (\ref{CLOG-2}). Similarly, we can derive that
\begin{eqnarray}\label{A-3}
&&logit\left(P\{Y(x^*,\mathbf{M}(x^*))=1|\mathbf{Z}\}\right)\nonumber\\
&&=c+ \gamma x^*  + \bm{\theta}^\prime \mathbf{Z} + \bbeta^\prime \tilde{\mathbf{c}} + \balpha^\prime \bbeta x^* + \sum_{k=1}^d \beta_k{\boldsymbol\eta}_k^\prime \mathbf{Z}+\frac{1}{2}\bbeta^\prime {\boldsymbol\Sigma}_e \bbeta,
\end{eqnarray}
and
\begin{eqnarray}\label{A-4}
&&logit\left(P\{Y(x,\mathbf{M}(x))=1|\mathbf{Z}\}\right)\nonumber\\
&&=c+ \gamma x  + \bm{\theta}^\prime \mathbf{Z} + \bbeta^\prime \tilde{\mathbf{c}} + \balpha^\prime \bbeta x + \sum_{k=1}^d \beta_k{\boldsymbol\eta}_k^\prime \mathbf{Z}+\frac{1}{2}\bbeta^\prime {\boldsymbol\Sigma}_e \bbeta.
\end{eqnarray}

Based on (\ref{A-1}), (\ref{A-2}) and (\ref{A-3}),  the calculation of the following expression is straightforward,
\begin{align*}
\log\left(\rm {NDE^{OR}}\right)= \gamma (x-x^*).
\end{align*}
Therefore, by taking exponential arithmetic we have
\begin{eqnarray*}
\rm {NDE^{OR}}&=& \exp\{\gamma(x-x^*)\}.
\end{eqnarray*}
In addition, (\ref{A-2}), (\ref{A-3}) and (\ref{A-4}) lead to that
\begin{eqnarray}\label{A-5}
\log\left(\rm {NIE^{OR}}\right)&=&\log\left\{\frac{P\{Y(x,\mathbf{M}(x))=1|\mathbf{Z}\}/[1-P\{Y(x,\mathbf{M}(x))=1|\mathbf{Z}\}]}{P\{Y(x,\mathbf{M}(x^*))=1|
\mathbf{Z}\}/[1-P\{Y(x,\mathbf{M}(x^*))=1|\mathbf{Z}\}]}\right\}\\
&=& logit\left(P\{Y(x,\mathbf{M}(x))=1|\mathbf{Z}\}\right) - logit\left(P\{Y(x,\mathbf{M}(x^*))=1|\mathbf{Z}\}\right)\nonumber\\
&=&\balpha^\prime \bbeta (x-x^*),\nonumber
\end{eqnarray}
and
\begin{eqnarray}\label{A-6}
\log\left(\rm {TE^{OR}}\right)&=&\log\left\{\frac{P\{Y(x,\mathbf{M}(x))=1|\mathbf{Z}\}/[1-P\{Y(x,\mathbf{M}(x))=1|\mathbf{Z}\}]}{P\{Y(x^*,\mathbf{M}(x^*))=1|
\mathbf{Z}\}/[1-P\{Y(x^*,\mathbf{M}(x^*))=1|\mathbf{Z}\}]}\right\}\\
&=& logit\left(P\{Y(x,\mathbf{M}(x))=1|\mathbf{Z}\}\right) - logit\left(P\{Y(x^*,\mathbf{M}(x^*))=1|\mathbf{Z}\}\right)\nonumber\\
&=&(\gamma + \balpha^\prime \bbeta) (x-x^*).\nonumber
\end{eqnarray}

Taking exponential arithmetic for both (\ref{A-5}) and (\ref{A-6}), we can get $\rm {NIE^{OR}} = \exp\{\balpha^\prime \bbeta (x-x^*)\}$ and $\rm {TE^{OR}} = \exp\{(\gamma + \balpha^\prime \bbeta) (x-x^*)\}$. This ends the proof.

}
\vspace{0.5cm}
%%%%%%%%%%%%%%%%%%%%%%5
\bibliographystyle{natbib}
\bibliography{reference}

%%%%%%%%%%%%%%%%%%%%55
%%%%%%%%%%%%%%%%%%%%%%%%%%%%%%%%%%%%%%%%%%%%%%%%%%%%%%%%%%%%%%%%%%%%%%%%%%%%%%%%%%%%%%%%%%%%%%%%%%%%%%%%%%%%%

%%%%%%%%%%%%%%%%%%%%%%%%%%%

\end{document}